# Identification of coherent twin relationship from high-resolution reciprocal space maps

Authors

**Semën Gorfman[a]\*, David Spirito[a], Guanjie Zhang[b], Carsten Detlefs[c] and Nan Zhang[b]**

[a]Department of Materials Science and Engineering, Tel Aviv University, Wolfson Building for Mechanical Engineering, Tel Aviv, 6997801, Israel

[b]Electronic Materials Research Laboratory, Key Laboratory of the Ministry of Education and International Center for Dielectric Research, School of Electronic and Information Engineering, Xi'an Jiaotong University, Xi'an, People's Republic of China

[c] European Synchrotron Radiation Facility, 6, rue Jules Horowitz, Grenoble, 38043, France

Correspondence email: gorfman@tauex.tau.ac.il

**Funding information**    Israel Science Foundation (grant No. 1561/18; award No. 2247/18 (equipment grant)); US-Israel Binational Science Foundation (grant No. 2018161).

**Synopsis**   We present the theory and an algorithm for the assignment of ferroelastic domains to the individual components of split Bragg peaks in high-resolution reciprocal space maps. We further develop the formalism of mechanical compatibility of ferroelastic domains for the analysis of the geometry of the reciprocal space. The application of the algorithm to the reciprocal space maps of tetragonal $BaTiO_3$ and rhombohedral $PbZr_{1-x}Ti_xO_3$ crystals is demonstrated.

**Abstract**    Twinning is a common crystallographic phenomenon, which usually occurs in crystals during symmetry-lowering phase transition. Once formed, twin domains play an important role in defining physical properties: for example, twin domains underpin the giant piezoelectric effect in ferroelectrics, superelasticity in ferroelastics and the shape-memory effect in martensitic alloys. Unfortunately, there is still a lack of experimental methods for imaging and characterization of twin domain patterns. Here, we propose a theoretical framework and an algorithm for the recognition of twinned pairs of ferroelastic domains and the identification of the coherent twin relationship using high-resolution reciprocal space mapping of X-ray diffraction intensity around split Bragg peaks. Specifically, we adapt the geometrical theory of twinned ferroelastic crystals (Fousek & Janovec, 1969) for the analysis of the X-ray diffraction patterns. We derive the necessary equations and outline an algorithm for calculation of the separation between the Bragg peaks, diffracted from possible coherent twin domains, connected to one another via mismatch-free interface. We demonstrate that such separation is always perpendicular to the planar interface between mechanically matched domains. As examples, we present the analysis of the separation between the peaks diffracted from tetragonal and rhombohedral domains in the high-resolution reciprocal space maps of $BaTiO_3$ and $PbZr_{1-x}Ti_xO_3$ crystals. The demonstrated method can be used to analyse the response of multi-domain patterns to external perturbations such as electric field, change of a temperature or pressure.

**Keywords: Ferroelastic domains, domain walls, High-resolution X-ray diffraction.**

1. Introduction

Twinning is a common crystallographic phenomena (Cahn, 1954; Grimmer & Nespolo, 2006; Authier, 2003). The presence of twin domains may alter or even dominate materials properties (Seidel, 2012; Catalan *et al.*, 2012; Tagantsev *et al.*, 2010), especially when a twin domain hosts multiple order parameters of different physical nature (e.g. electric polarization and mechanical strain). Domain switching, domain rearrangement and domain-wall motion may underpin / enhance the technologically important piezoelectric effect (Hu *et al.*, 2020), dielectric permittivity (Damjanovic, 1999; Trolier-McKinstry *et al.*, 2018), superelasticity (Viehland & Salje, 2014), the shape-memory effect (Bhattacharya, 2003) and domain-wall superconductivity (Catalan *et al.*, 2012). The knowledge of domain patterns (e.g. average domain sizes and shapes, domain-wall orientation) is important for materials design and properties engineering. More generally, twinning is the subject of much interest in fundamental and applied science.

Unfortunately, only a handful of experimental techniques are available for the experimental characterization of domain patterns (Wu *et al.*, 2015). These techniques are based on e.g. optical and birefringence microscopy (Ushakov *et al.*, 2019; Gorfman *et al.*, 2012), or piezo response-force (PFM) (Gruverman *et al.*, 2019) microscopies or X-ray topography (Yamada *et al.*, 1966). Each of these techniques have some disadvantages, limiting the completeness and efficiency of the characterization. For example, optical microscopy is almost insensitive to the strain / lattice parameters, PFM is only limited to the surface. Accordingly, any new way of characterization of domains in the bulk would contribute to the subject greatly.

Single crystal X-ray diffraction could potentially fill this methodological gap. It is bulk penetrating, it is non-destructive and it has structural characterization power (e.g. sensitivity to the lattice parameters). Using synchrotron radiation adds the capabilities for in-situ (e.g. stroboscopic) studies of domains at variable temperature and external electric field (see e.g. (Zhang *et al.*, 2018; Gorfman *et al.*, 2020)). The development of dark-field X-ray microscopy (Poulsen *et al.*, 2017; Kutsal *et al.*, 2019; Simons *et al.*, 2015) and coherent Bragg diffraction imaging methods (Robinson & Miao, 2004; Marçal *et al.*, 2020; Dzhigaev *et al.*, 2021) for combining reciprocal and real space information is another step towards advanced characterization of domain patterns. However, despite the great potential of X-ray diffraction for the characterization of domain patterns, the technique remains far from routine. It is mainly because the geometry of X-ray diffraction from a multi-domain crystal may be as complex as domain patterns themselves.

We propose a framework for the recognition of coherent twin relationship using high-resolution three-dimensional reciprocal space mapping. Here the word "coherent" describes the situation when two (or more) domains alternate and connect to one-another without a lattice mismatch. If formed and stable, such twin domain patterns may significantly enhance the ability of a material to respond to external perturbations and thus enable new domain-related physical properties. Specifically we focus on the assemblies of ferroelastic domains (the volumes of a crystal where strain is uniform). While the method assumes the availability of the reciprocal space information alone, it may also assist in the interpretation of the dark field X-ray microscopy data.

The manuscript has the following organization. After introducing the glossary of symbols and important relationship, we recapitulate the well-known formalism for mechanical compatibility of coherent patterns of ferroelastic domains in a way that is suitable for the analysis of X-ray diffraction from them. Then we analyse the orientation relationship between their reciprocal lattices and calculate reciprocal space separation of Bragg peaks of twinned ferroelectric domains. The demonstration of the method for the identification of coherent twin relationship in domains of tetragonal ($BaTiO_3$) and rhombohedral ($PbZr_{0.75}Ti_{0.25}O_3$) symmetry is presented. We inspect diffraction from multi-domain ferroelectric crystals accordingly and show the way to assign different peak components to the individual domains.

**2. Glossary of symbols and important relationship**

The goal of this chapter is to introduce the important notations and relationship. We do not define the reasons for these introductions at this point, but note that they enable the descriptions of all the necessary real and reciprocal space phenomena in the shortest way.

<u>Basis vectors:</u> $\boldsymbol{a}_{im}$ ($i = 1..3$) are the basis vectors of a crystal lattice[1]. The second index refers to the ferroelastic domain variant $m$. $m = 0$ corresponds to the crystal lattice of higher-symmetry (e.g. cubic) "parent" phase (Figure 1). The parallelepiped based on the vectors $\boldsymbol{a}_{im}$ forms a unit cell.

<u>Unit cell settings:</u> Many unit cell settings exist for the same lattice (Gorfman, 2020). Here, we prefer the cell settings $\boldsymbol{a}_{im}$ ($m > 0$) obtained by the small distortion / rotation of the parent phase basis vectors $\boldsymbol{a}_{i0}$. Figure 1 shows two-dimensional illustration of two ferroelastic domains and the settings $\boldsymbol{a}_{i1}$, $\boldsymbol{a}_{i2}$ and $\boldsymbol{a}_{i0}$ for the domains (1), (2) and (0).

---

[1] The terms lattice and structure are often misused in the recent materials science literature (as noticed by (Nespolo, 2019)). Therefore, we underline that "a crystal lattice" refers to a regular array of points accounting for the periodicity of the structure. On the contrary, "a crystal structure" is obtained by translating a unit cell to all the points of a crystal lattice.

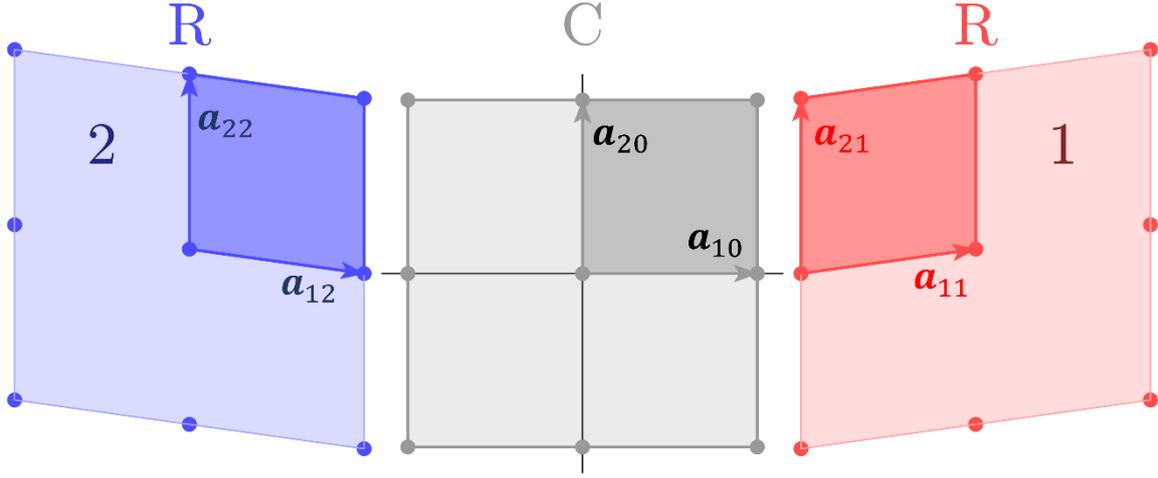

**Figure 1** Schematic illustration of two-dimensional ferroelastic domains and their unit cells (2x2 supercells are shown for clarity). The middle image (marked by the letter C, standing for the two-dimensional prototype of "cubic") corresponds to the single-domain "parent" phase. The lattice basis vectors here are $a_{i0}$. The right and left images (marked by the letter R, standing for the two-dimensional prototype of "rhombohedral") correspond to the ferroelastic domains. The lattice basis vectors here ($a_{im}$, $m = 1 \ldots 2$) are chosen in such a way that $a_{im}$ are nearly parallel to $a_{i0}$.

<u>Metric tensor / matrix of dot products:</u> $G_{ij} = a_i a_j$ is the metric tensor (Giacovazzo, 1992; Hahn, 2005). The corresponding 3x3 matrix $[G]_m$ is the matrix of dot products for the domain variant $m$. Their determinants are $|G| = V_A^2$ ($V_A$ is the unit cell volume). For a cubic lattice, $G_{ij} = a_0^2 \delta_{ij}$ is valid (here $a_0$ is the "cubic" lattice parameter and $\delta_{ij}$ is the Kronecker symbol).

<u>Cartesian / crystal physical coordinate system:</u> $e_i$ are the basis vectors, such that $e_i e_j = \delta_{ij}$. $e_i$ are fixed with respect to $a_{i0}$ (e.g. $a_{i0} = a_0 e_i$ for the cubic lattice).

<u>Parent phase twinning matrix:</u> $[T]$ represents a symmetry operation of the parent phase lattice (basis vectors $a_{i0}$) that is not such of a ferroelastic phase lattice. We define $[T]$ as 3x3 matrix, which describes the transformation of the Cartesian coordinate system $e_i$ using following formal matrix equation:

$$(e_1' \quad e_2' \quad e_3') = (e_1 \quad e_2 \quad e_3) \begin{pmatrix} T_{11} & T_{12} & T_{13} \\ T_{21} & T_{22} & T_{23} \\ T_{31} & T_{32} & T_{33} \end{pmatrix}, \qquad (1)$$

Note that the matrix $[T]$ is orthogonal ($[T]^{-1} = [T]^T$). The following relationship between the matrices of dot products of ferroelastic domains ($m$) and (1) holds (see Appendix A for the proof):

$$[G]_m = [T]_m [G]_1 [T]_m^T \qquad (2)$$

The two-dimensional example in the Figure 1, shows ferroelastic domains (1) and (2). The metric tensor of the domain (1) is $[G]_1 = a^2 \begin{pmatrix} 1 & \eta \\ \eta & 1 \end{pmatrix}$ (here $\eta$ is the cosine of the angle between the basis vectors).

Domain (2) is related to the domain (1) by twinning matrix $[T] = \begin{pmatrix} \bar{1} & 0 \\ 0 & 1 \end{pmatrix}$ so that $[G]_2 = a^2 \begin{pmatrix} \bar{1} & 0 \\ 0 & 1 \end{pmatrix} \begin{pmatrix} 1 & \eta \\ \eta & 1 \end{pmatrix} \begin{pmatrix} \bar{1} & 0 \\ 0 & 1 \end{pmatrix} = a^2 \begin{pmatrix} 1 & \bar{\eta} \\ \bar{\eta} & 1 \end{pmatrix}$.

The following notations are introduced for the compactness of the description of the connectivity of lattices of domains $n$ and $m$.

$[\Delta G] = [G]_n - [G]_m$ is the difference between the metric tensors of the domain $n$ and $m$. The reference to domain numbers in $[\Delta G]$ is omitted below.

<u>Eigenvalues and eigenvectors:</u> $\lambda_1, \lambda_2, \lambda_3$ are the eigenvalues of $[\Delta G]$, $[V]$ is the matrix of the corresponding eigenvectors (defining the columns of the matrix). The condition $\Delta G_{ij} = \Delta G_{ji}$ means that $\lambda_i$ are real and that $[V]$ is orthogonal ($[V]^{-1} = [V]^T$). $V_{ij}$ are the elements of the matrix $[V]$.

<u>Supplementary coordinate systems:</u> the set of vectors $\boldsymbol{v}_i, \boldsymbol{u}_i$ and $\boldsymbol{w}_i$ define the following coordinate systems.

$$(\boldsymbol{v}_1 \quad \boldsymbol{v}_2 \quad \boldsymbol{v}_3) = (\boldsymbol{a}_1 \quad \boldsymbol{a}_2 \quad \boldsymbol{a}_3) \begin{pmatrix} V_{11} & V_{12} & V_{13} \\ V_{21} & V_{22} & V_{23} \\ V_{31} & V_{32} & V_{33} \end{pmatrix} \quad (3)$$

$\boldsymbol{u}_i$ are introduced when $\lambda^{(1,3)} \neq 0$:

$$(\boldsymbol{u}_1 \quad \boldsymbol{u}_2 \quad \boldsymbol{u}_3) = (\boldsymbol{v}_1 \quad \boldsymbol{v}_2 \quad \boldsymbol{v}_3)[\Lambda], \quad [\Lambda] = \begin{bmatrix} 1 & 0 & 0 \\ 0 & 1 & 0 \\ 0 & 0 & \sqrt{\frac{|\lambda_1|}{|\lambda_3|}} \end{bmatrix} \quad (4)$$

$\boldsymbol{w}_i$ are introduced together with $\boldsymbol{u}_i$:

$$(\boldsymbol{w}_1 \quad \boldsymbol{w}_2 \quad \boldsymbol{w}_3) = (\boldsymbol{u}_1 \quad \boldsymbol{u}_2 \quad \boldsymbol{u}_3)[Z], \quad [Z] = \begin{bmatrix} 1 & 0 & 0 \\ 0 & 1 & 0 \\ \pm 1 & 0 & 1 \end{bmatrix} \quad (5)$$

The transformation between $\boldsymbol{a}_i$ to $\boldsymbol{w}_i$ is achieved by the matrix

$$[W] = [V][\Lambda][Z]. \quad (6)$$

<u>Vector coordinates:</u> $x_i, u_i, v_i, w_i$ are the coordinates of an arbitrary vector with respect to the coordinate systems $\boldsymbol{a}_i, \boldsymbol{v}_i, \boldsymbol{u}_i$ and $\boldsymbol{w}_i$, so that $x_i \boldsymbol{a}_i = u_i \boldsymbol{u}_i = v_i \boldsymbol{v}_i = w_i \boldsymbol{w}_i$. The following direct and inverse transformation between the coordinates apply:

$$\begin{aligned} x_i &= V_{ij} v_j \\ v_i &= V_{ji} x_j \end{aligned} \quad (7)$$

and

$$v_{1,2} = u_{1,2} \quad v_3 = \sqrt{\frac{|\lambda_1|}{|\lambda_3|}} u_3 \quad (8)$$

$$u_{1,2} = w_{1,2} \quad u_3 = Z_{31} w_1 + w_3$$

Supplementary metric tensors / matrices of dot products: $[G^{(V)}], [G^{(U)}], [G^{(W)}]$ are defined by the dot products $G_{Vij} = \boldsymbol{v}_i \boldsymbol{v}_j$, $G_{Uij} = \boldsymbol{u}_i \boldsymbol{u}_j$, $G_{Wij} = \boldsymbol{w}_i \boldsymbol{w}_j$. The following relationship apply (here $X \equiv U, V, W$):

$$[G^{(X)}] = [X]^T [G][X] \tag{9}$$

The transformation matrix: the transformation e.g. between the basis vectors $\boldsymbol{a}_{im}$ and $\boldsymbol{a}_{in}$ is defined by the transformation matrix $[S]$. The columns of the matrix $[S]$ are the coordinates of $\boldsymbol{a}_{1n}, \boldsymbol{a}_{2n}, \boldsymbol{a}_{3n}$ with respect to $\boldsymbol{a}_{1m}, \boldsymbol{a}_{2m}, \boldsymbol{a}_{3m}$.

$$(\boldsymbol{a}_{1n} \quad \boldsymbol{a}_{2n} \quad \boldsymbol{a}_{3n}) = (\boldsymbol{a}_{1m} \quad \boldsymbol{a}_{2m} \quad \boldsymbol{a}_{3m}) \begin{pmatrix} S_{11} & S_{12} & S_{13} \\ S_{21} & S_{22} & S_{23} \\ S_{31} & S_{32} & S_{33} \end{pmatrix} \tag{10}$$

Reciprocal basis vectors: The superscript * refers to the reciprocal bases, e.g. $\boldsymbol{a}_i^*$ or $\boldsymbol{w}_i^*$ are such that $\boldsymbol{a}_i \boldsymbol{a}_j^* = \boldsymbol{w}_i \boldsymbol{w}_j^* = \delta_{ij}$.

Transformation between the reciprocal basis vectors: If the direct basis vectors (e.g. $\boldsymbol{a}_{im}$ and $\boldsymbol{a}_{in}$) are related by the matrix $[S]$ (according to the equation (10)) then the corresponding reciprocal lattice vectors $\boldsymbol{a}_{im}^*$ and $\boldsymbol{a}_{in}^*$) are related by the matrix $[S^*]$. The following relationship between $[S]$ and $[S^*]$ holds:

$$[S^*]^T = [S]^{-1} \tag{11}$$

Reciprocal coordinates of a vector: $x_i^*, w_i^*$ are the coordinates of an arbitrary vector with respect to the reciprocal coordinate system $x_i^* \boldsymbol{a}_i^* = w_i^* \boldsymbol{w}_i^*$. The vector, indicating the position in the reciprocal space / lattice is denoted by $\boldsymbol{B}$. We also the notations, $h\ k\ l$ for the indices of a plane and $H\ K\ L$ for the indices of Bragg reflection. In the case if the direct basis vectors (e.g. $\boldsymbol{a}_i$ and $\boldsymbol{w}_i$) are related by the orientation matrix $[W]$ then the transformation between the corresponding reciprocal lattice coordinates are given by (according to the equations (7) and (11)) by

$$w_i^* = W_{ji} x_j^* \tag{12}$$

## 3. Mismatch-free connection of domains

This paragraph recapitulates the approach of (Fousek & Janovec, 1969; Sapriel, 1975) for the description of the geometrical connectivity of ferroelastic domains. Importantly, it disregards the connectivity of atoms (e.g. oxygen octahedra in perovskites (Beanland, 2011)) but rather considers connectivity of lattices only. The lattices $(n)$ and $(m)$ are considered as connected if they meet along their common $(hkl)$ plane such that those have exactly the same in-$(hkl)$-plane two-dimensional lattice parameters. The theory of martensitic phase transformations (Bhattacharya, 2003) refers to such planes as to "habit planes". All the points in this plane should have coordinates $x_i$ such that:

$$\Delta G_{ij} x_i x_j = 0 \tag{13}$$

We are searching for the cases when (13) can be reduced to the equation of a plane:

$$hx_1 + kx_2 + lx_3 = 0 \tag{14}$$

Let us transform the coordinates $x_i$ to $v_i$ according to (7) (and so the coordinate system $a_i$ to $v_i$ according to (3)). This will simplify (13) to

$$\lambda_1 v_1^2 + \lambda_2 v_2^2 + \lambda_3 v_3^2 = 0 \tag{15}$$

The equation (15) can be re-written as (14) if e.g. $\lambda_2 = 0$. Two cases may be considered:

The case $\lambda_1 = \lambda_2 = 0$, $\lambda_3 \neq 0$ yields:

$$v_3 = 0, \tag{16}$$

and represent $(001)_v$ plane (the subscript $v$ refers to the Miller indices with respect to $v_i$ instead of $a_i$). Using (7), we reformulate (16) as

$$V_{13}x_1 + V_{23}x_2 + V_{33}x_3 = 0 \tag{17}$$

Extending the components $V_{13}, V_{23}, V_{33}$ to the integer numbers will give the Miller indices $h\,k\,l$ of the mismatch-free plane.

The case $\lambda_1 < 0, \lambda_2 = 0, \lambda_3 > 0$ leads to two possible plane solutions of (15):

$$|\lambda_1|^{\frac{1}{2}} v_1 \pm |\lambda_3|^{\frac{1}{2}} v_3 = 0 \tag{18}$$

Let us transform the coordinates $v_i$ to $u_i$ (or $w_i$) using (8). This will re-write (18) as

$$u_1 \pm u_3 = 0, \text{ or equivalently}$$
$$w_3 = 0 \tag{19}$$

These are the equations for $(101)_u$ and $(10\bar{1})_u$ planes correspondingly (the subscript $u$ refers to the Miller indices with respect to $u_i$). Alternatively, (19) can be described as $(001)_w$ ($Z_{31} = 1$ is taken for $(10\bar{1})_u$ and $Z_{31} = -1$ is taken for $(101)_u$ planes). Using (4) and (8) can help rewriting (18) as:

$$(\sqrt{|\lambda_1|}V_{i1} \pm \sqrt{|\lambda_3|}V_{i3})x_i = 0 \tag{20}$$

Extending the components of $(\sqrt{|\lambda_1|}V_{i1} \pm \sqrt{|\lambda_3|}V_{i3})$ to the integer numbers would give the Miller indices $h\,k\,l$ of the habit planes between the domains. This formalism gives well-known results for the Miller indices of the possible habit planes between the domains of different symmetry. Some examples of such are presented further in the paragraphs 6 and 7.

### 4. Mutual orientation of domains

The goal of this paragraph is to find the transformation matrix $[S]$ (as defined by (10)) between the basis vectors $a_{im}$ and $a_{in}$ when the lattices of the domains $m$ and $n$ meet along their common $(001)_w$ plane. Let us first find the similar transformation matrix $[S_w]$ between $w_{im}$ and $w_{in}$. Considering the matching along $(001)_w$ plane implies $w_{1,2m} = w_{1,2n}$ so that:

$$[S_w] = \begin{pmatrix} 1 & 0 & y_1 \\ 0 & 1 & y_2 \\ 0 & 0 & y_3 \end{pmatrix} \tag{21}$$

The unknown coefficients $y_i$ can be calculated as follows. First, consider that the determinant of the transformation matrix ($|S_w| = y_3$) should be equal to the ratio of the unit cell volumes, therefore

$$y_3 = \sqrt{\frac{\det([G]_n)}{\det([G]_m)}} \tag{22}$$

Note that, although $y_3 = 1$ for the same phase domains, the formalism is valid for the description of the matching between the lattices of different symmetry and unit cell volume ($y_3 \neq 1$). $y_1, y_2$ can be found using (9) ($[G^{(W)}]_n = [S_w]^T [G^{(W)}]_m [S_w]$) substituting (21) into (9) yields:

$$[G^{(W)}]_n = \begin{bmatrix} G^{(W)}_{11,m} & G^{(W)}_{12,m} & G^{(W)}_{1i,m} y_i \\ G^{(W)}_{12,m} & G^{(W)}_{22,m} & G^{(W)}_{2i,m} y_i \\ G^{(W)}_{i1,m} y_i & G^{(W)}_{i2,m} y_i & G^{(W)}_{ij,m} y_i y_j \end{bmatrix} \tag{23}$$

Comparing the elements $G^{(W)}_{13}$ and $G^{(W)}_{23}$ of the matrices, we get:

$$\begin{cases} G^{(W)}_{11,m} y_1 + G^{(W)}_{12,m} y_2 = G^{(W)}_{13,n} - G^{(W)}_{13,m} y_3 \\ G^{(W)}_{21,m} y_1 + G^{(W)}_{22,m} y_2 = G^{(W)}_{23,n} - G^{(W)}_{23,m} y_3 \end{cases} \tag{24}$$

Solving this system of linear equations gives the values of the remaining coefficients $y_1, y_2$ and accordingly all the elements of the matrix $[S_w]$. Finally, the $[S]$ can be found according to the equation:

$$(\boldsymbol{a}_{1n} \quad \boldsymbol{a}_{2n} \quad \boldsymbol{a}_{3n})[W] = (\boldsymbol{a}_{1m} \quad \boldsymbol{a}_{2m} \quad \boldsymbol{a}_{3m})[W][S_w] \tag{25}$$

Which immediately leads to

$$[S] = [W][S_w][W]^{-1} \tag{26}$$

Accordingly, the matrix $[S]$ can be found by going through the following steps. The corresponding numerical examples will be presented in the §6 and §7.

- Choosing appropriate twinning matrices, $[T]_m$ and $[T]_n$ and calculation the elements of the corresponding metric tensors $[G]_m$ and $[G]_n$ using (2).
- Calculating the eigenvectors and eigenvalues of $[\Delta G] = [G]_n - [G]_m$.
- Using these eigenvalues and eigenvectors to form the matrices $[V], [\Lambda], [Z]$ and $[W]$ according to the equations (3), (4), (5), (6). For the cases when two eigenvalues of $[\Delta G]$ are zero, $[Z]$ is a unitary matrix.
- Calculating $[G^{(W)}]_m$ and $[G^{(W)}]_n$ according to the equation (9) and determining the coefficients $y_1, y_2, y_3$ using (22) and (24). Setting the matrix $[S_w]$ according to (21) and converting it to $[S]$ according to (26).

## 5. Separation between the Bragg peaks

Different twin domains would diffract X-rays into slightly different directions. The corresponding nods of the reciprocal lattices can nearly overlap in some cases but be resolved in others. The examples of real diffraction patterns of perovskite-based crystals with ferroelastic domains can be seen in (Gorfman & Thomas, 2010; Gorfman *et al.*, 2011; Choe *et al.*, 2018; Zhang *et al.*, 2018; Gorfman *et al.*, 2020). Figure 2a shows a two-dimensional example of domains, matching along $(1\bar{1})$ planes. Figure 2b shows their reciprocal lattices. The goal of this paragraph is to calculate the separation $\Delta \boldsymbol{B}$ between the Bragg peaks $H\ K\ L$ of two matched domain variants $(m)$ and $(n)$. Specifically we will derive the coordinates $\Delta H\ \Delta K\ \Delta L$ of $\Delta \boldsymbol{B}$ relative to the reciprocal basis vectors of the domain $m$ ($\boldsymbol{a}^*_{im}$). Let us first calculate the $\Delta H_w\ \Delta K_w\ \Delta L_w$ coordinates of $\Delta \boldsymbol{B}$ with respect to $\boldsymbol{w}^*_{im}$. We express $\Delta \boldsymbol{B}$ in the form:

$$\Delta \boldsymbol{B} = (\Delta \boldsymbol{w}^*_1 \quad \Delta \boldsymbol{w}^*_2 \quad \Delta \boldsymbol{w}^*_3)\begin{pmatrix} H_w \\ K_w \\ L_w \end{pmatrix}, \quad \Delta \boldsymbol{w}^*_i = \boldsymbol{w}^*_{in} - \boldsymbol{w}^*_{im} \ . \tag{27}$$

Considering that $\boldsymbol{w}^*_{in}$ and $\boldsymbol{w}^*_{im}$ are related by the matrix $[S^*_w]$ we get:

$$\Delta \boldsymbol{B} = (\boldsymbol{w}^*_{1m} \quad \boldsymbol{w}^*_{2m} \quad \boldsymbol{w}^*_{3m})([S^*_w] - [I])\begin{pmatrix} H_w \\ K_w \\ L_w \end{pmatrix} \tag{28}$$

So that

$$\begin{pmatrix} \Delta H_w \\ \Delta K_w \\ \Delta L_w \end{pmatrix} = ([S^*_w] - [I])\begin{pmatrix} H_w \\ K_w \\ L_w \end{pmatrix} \tag{29}$$

Using (21) and considering that $[S^*_w]^T = [S_w]^{-1}$ we get:

$$[S^*_w] = y_3^{-1}\begin{pmatrix} 1 & 0 & 0 \\ 0 & 1 & 0 \\ -y_1 & -y_2 & 1 \end{pmatrix}, \quad [S^*_w] - [I] = y_3^{-1}\begin{pmatrix} 0 & 0 & 0 \\ 0 & 0 & 0 \\ -y_1 & -y_2 & 1 - y_3 \end{pmatrix} \tag{30}$$

Substituting (30) to (29) we obtain

$$\begin{aligned} \Delta H_w &= 0 \\ \Delta K_w &= 0 \\ \Delta L_w &= -\frac{y_1}{y_3}H_w - \frac{y_2}{y_3}K_w + \frac{1-y_3}{y_3}L_w \end{aligned} \tag{31}$$

This also means that

$$\Delta \boldsymbol{B} = \left(-\frac{y_1}{y_3}H_w - \frac{y_2}{y_3}K_w + \frac{1-y_3}{y_3}L_w\right)\boldsymbol{w}^{*(m)}_3 \tag{32}$$

so that $\Delta \boldsymbol{B} \parallel \boldsymbol{w}^*_{3m}$. Considering that $\boldsymbol{w}^*_{3m} \perp (001)_w$, we conclude that $\Delta \boldsymbol{B}$ is normal to the domain wall. This statement is graphically illustrated in the Figure 2, which shows two lattices matched along $(1\bar{1})$ plane. Note that expression (29) can be reformulated in order to express the separation vector relative to the coordinate system $\boldsymbol{a}^*_{im}$:

$$\begin{pmatrix} \Delta H \\ \Delta K \\ \Delta L \end{pmatrix} = ([S^*] - [I])\begin{pmatrix} H \\ K \\ L \end{pmatrix} \tag{33}$$

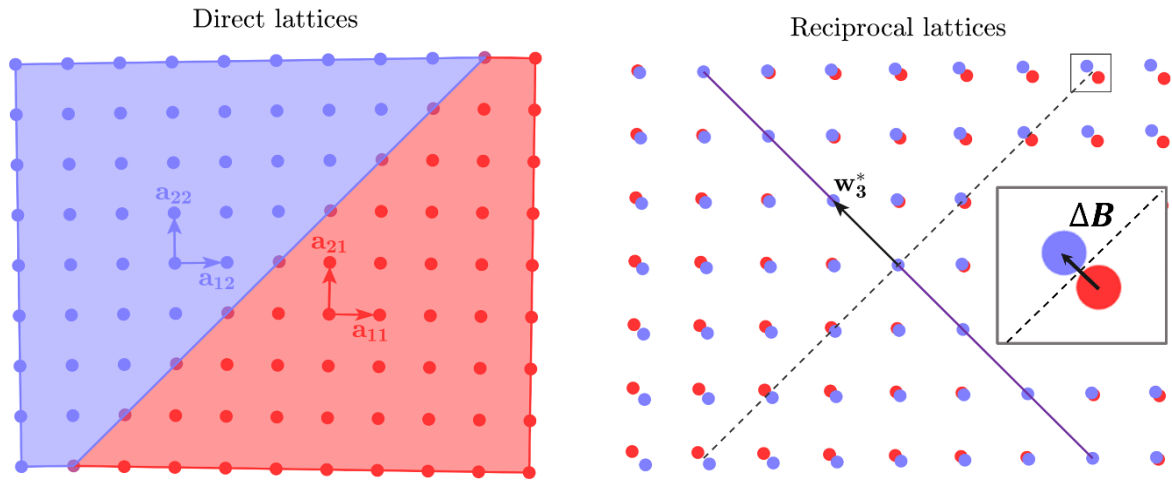

**Figure 2** Two-dimensional illustration of direct and reciprocal lattices of two domains. (a) The lattices of two 2D-tetragonal (=rectangular) domains connected along their common $(1\bar{1})$ plane. (b) Their corresponding reciprocal lattices. The dashed line is parallel to the $(1\bar{1})$ plane (domain wall), the inset highlights the separation between corresponding reciprocal lattice vectors, showing that it is perpendicular to the domain wall.

The next two chapters demonstrate the formalism on the examples of domains of tetragonal and rhombohedral symmetry.

## 6. Examples

### 6.1. Tetragonal domains

Let us assume that paraelastic / ferroelastic phases belong to the point symmetry groups $m3m / \frac{4}{m}mm$. Because these groups contain 48 and 16 symmetry operations respectively, the phase transition between them results in the formation of three domain variants (Figure 3 and Table 1).

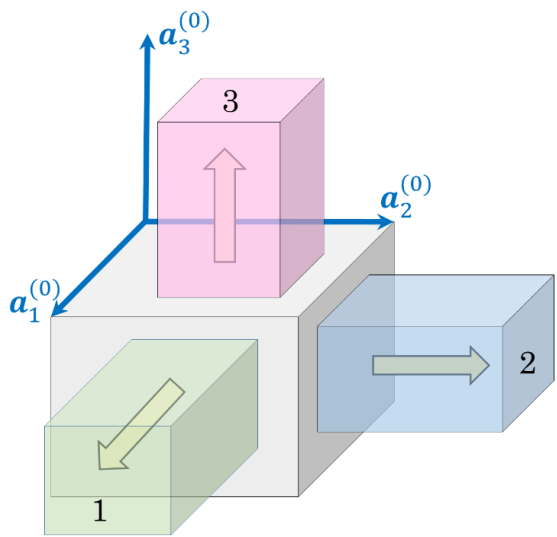

1 ($a$ domain): $\boldsymbol{P} \parallel [100]$

2 ($b$ domain): $\boldsymbol{P} \parallel [010]$

3 ($c$ domain): $\boldsymbol{P} \parallel [001]$

**Figure 3** Definition and numbering of the tetragonal domain variants. The allowed direction of the spontaneous polarization are given with respect to the basis vectors of the domains $\boldsymbol{a}_{im}$ ($m = 1..3$).

**Table 1** The definition of tetragonal domain variants (Figure 3). The first row assigns the number (name) to the domain variant; the second row shows the matrix of dot products $[G]_m$; the third row shows crystallographic direction (e.g. with respect to the lattice basis vectors $\boldsymbol{a}_{im}$) of the spontaneous polarization.

| Domain number (name) | 1 ($a$) | 2 ($b$) | 3 ($c$) |
| --- | --- | --- | --- |
| Metric tensor $[G]_m$ | $\begin{pmatrix} c^2 & 0 & 0 \\ 0 & a^2 & 0 \\ 0 & 0 & a^2 \end{pmatrix}$ | $\begin{pmatrix} a^2 & 0 & 0 \\ 0 & c^2 & 0 \\ 0 & 0 & a^2 \end{pmatrix}$ | $\begin{pmatrix} a^2 & 0 & 0 \\ 0 & a^2 & 0 \\ 0 & 0 & c^2 \end{pmatrix}$ |
| Spontaneous polarization direction $[P]_m$ | [1 0 0] | [0 1 0] | [0 0 1] |

Consider the connectivity between the domains $1(a)$ and $3(c)$. Following §3 we get

$$[\Delta G] = \begin{pmatrix} a^2 - c^2 & 0 & 0 \\ 0 & 0 & 0 \\ 0 & 0 & c^2 - a^2 \end{pmatrix} \tag{34}$$

$$\lambda_2 = 0, \quad \lambda_1 = -\lambda_3 = a^2 - c^2$$

Because $[\Delta G]$ is diagonal and $\lambda_2 = 0$ we can set $[V] = [I]$ and immediately obtain that (according to the equations (18), (19), (20)) domains may match along $(10\bar{1})$ or $(101)$ planes.

**For the case of $(10\bar{1})$ domain wall:**

$$[W] = [Z] = \begin{pmatrix} 1 & 0 & 0 \\ 0 & 1 & 0 \\ 1 & 0 & 1 \end{pmatrix} \tag{35}$$

To formulate the system of equations (24) we need to calculate the matrices $[G^{(W)}]_{1,3} = [W]^T [G]_{1,3} [W]$. Using (35):

$$[G^{(W)}]_1 = \begin{pmatrix} c^2 + a^2 & 0 & a^2 \\ 0 & a^2 & 0 \\ a^2 & 0 & a^2 \end{pmatrix}, \quad [G^{(W)}]_3 = \begin{pmatrix} c^2 + a^2 & 0 & c^2 \\ 0 & a^2 & 0 \\ c^2 & 0 & c^2 \end{pmatrix} \tag{36}$$

We now have to find the unknown coefficients $y_1, y_2, y_3$. According to (22) $y_3 = 1$. The system of equations (24) can be re-written as

$$\begin{cases} y_1 = \tau \\ y_2 = 0 \end{cases} \tag{37}$$

Here we introduced the notation

$$\tau = \frac{c^2 - a^2}{c^2 + a^2} \tag{38}$$

Using (37) and (21) we obtain

$$[S_w] = \begin{pmatrix} 1 & 0 & \tau \\ 0 & 1 & 0 \\ 0 & 0 & 1 \end{pmatrix} \tag{39}$$

Now we substitute (35) and (39) to (26) and get the following equations for $[S]$ and $[S^*]$:

$$[S] = \begin{pmatrix} 1-\tau & 0 & \tau \\ 0 & 1 & 0 \\ -\tau & 0 & 1+\tau \end{pmatrix}, \quad [S^*] = \begin{pmatrix} 1+\tau & 0 & \tau \\ 0 & 1 & 0 \\ -\tau & 0 & 1-\tau \end{pmatrix} \tag{40}$$

and

$$[S^*] - [I] = \tau \begin{pmatrix} 1 & 0 & 1 \\ 0 & 0 & 0 \\ \bar{1} & 0 & \bar{1} \end{pmatrix} \tag{41}$$

This finally gives the following expression for the separation between $HKL$ Bragg peaks, diffracted from the domains $1(a)$ and $3(c)$:

$$\begin{pmatrix} \Delta H \\ \Delta K \\ \Delta L \end{pmatrix} = \tau(H+L) \begin{pmatrix} 1 \\ 0 \\ \bar{1} \end{pmatrix} \tag{42}$$

Identical analysis can be implemented to (101) domain wall and other pairs of domains. Table 2 summarizes the results: it includes all the possible mismatch-free domains walls and corresponding coordinates of $\Delta \boldsymbol{B}$ in the reciprocal coordinate system of the domain $m$. Note, that the separation vector $\Delta \boldsymbol{B}$ is always perpendicular to the domain wall.

**Table 2** The summary of possible mismatch-free domain walls between tetragonal domains and the corresponding separation between their Bragg peaks. The first two columns shows the domains numbers (names) $m$ and $n$ (according to Table 1 and Figure 3). The third column shows the Miller indices of the mismatch-free plane. The fourth column shows the matrix $[S]$ (the columns of this matrix are the coordinates of the vectors $\boldsymbol{a}_{ni}$ relative to the domain $\boldsymbol{a}_{mi}$). The fifth column shows the matrix $[S^*] - [I]$. The last column shows the coordinates of the vector $\Delta \boldsymbol{B} = \boldsymbol{B}_n - \boldsymbol{B}_m$ relative to the reciprocal basis vectors $\boldsymbol{a}^*_{mi}$.

| $m$ | $n$ | Plane | $[S]$ | $[S^*] - [I]$ | $\Delta \boldsymbol{B}$ |
|---|---|---|---|---|---|
| $1(a)$ | $2(b)$ | $(1\bar{1}0)$ | $\begin{pmatrix} 1-\tau & \tau & 0 \\ -\tau & 1+\tau & 0 \\ 0 & 0 & 1 \end{pmatrix}$ | $\begin{pmatrix} \tau & \tau & 0 \\ -\tau & -\tau & 0 \\ 0 & 0 & 1 \end{pmatrix}$ | $\tau(H+K)\begin{pmatrix} 1 \\ \bar{1} \\ 0 \end{pmatrix}$ |
| $1(a)$ | $2(b)$ | $(110)$ | $\begin{pmatrix} 1-\tau & -\tau & 0 \\ \tau & 1+\tau & 0 \\ 0 & 0 & 1 \end{pmatrix}$ | $\begin{pmatrix} \tau & -\tau & 0 \\ \tau & -\tau & 0 \\ 0 & 0 & 1 \end{pmatrix}$ | $\tau(H-K)\begin{pmatrix} 1 \\ 1 \\ 0 \end{pmatrix}$ |
| $1(a)$ | $3(c)$ | $(10\bar{1})$ | $\begin{pmatrix} 1-\tau & 0 & \tau \\ 0 & 1 & 0 \\ -\tau & 0 & 1+\tau \end{pmatrix}$ | $\begin{pmatrix} \tau & 0 & \tau \\ 0 & 1 & 0 \\ -\tau & 0 & -\tau \end{pmatrix}$ | $\tau(H+L)\begin{pmatrix} 1 \\ 0 \\ \bar{1} \end{pmatrix}$ |
| $1(a)$ | $3(c)$ | $(101)$ | $\begin{pmatrix} 1-\tau & 0 & -\tau \\ 0 & 1 & 0 \\ \tau & 0 & 1+\tau \end{pmatrix}$ | $\begin{pmatrix} \tau & 0 & -\tau \\ 0 & 1 & 0 \\ \tau & 0 & -\tau \end{pmatrix}$ | $\tau(H-L)\begin{pmatrix} 1 \\ 0 \\ 1 \end{pmatrix}$ |

| | | | | | | |
|---|---|---|---|---|---|---|
| 2(b) | 3(c) | (01$\bar{1}$) | $\begin{pmatrix} 1 & 0 & 0 \\ 0 & 1-\tau & \tau \\ 0 & -\tau & 1+\tau \end{pmatrix}$ | $\begin{pmatrix} 1 & 0 & 0 \\ 0 & \tau & \tau \\ 0 & -\tau & -\tau \end{pmatrix}$ | $\tau(K+L)$ | $\begin{pmatrix} 0 \\ 1 \\ \bar{1} \end{pmatrix}$ |
| 2(b) | 3(c) | (011) | $\begin{pmatrix} 1 & 0 & 0 \\ 0 & 1-\tau & -\tau \\ 0 & \tau & 1+\tau \end{pmatrix}$ | $\begin{pmatrix} 1 & 0 & 0 \\ 0 & \tau & -\tau \\ 0 & \tau & -\tau \end{pmatrix}$ | $\tau(K-L)$ | $\begin{pmatrix} 0 \\ 1 \\ 1 \end{pmatrix}$ |

## 6.2. Rhombohedral (trigonal) domains

Let us assume that paraelastic / ferroelastic phases belong to the cubic point symmetry group $m3m$ / $\bar{3}m$, containing 48 and 12 symmetry operations correspondingly. Accordingly, $m3m \rightarrow \bar{3}m$ transition results in the formation of four domain variants. These domains are illustrated in the in Figure 4 and Table 3. Analogously to the case of tetragonal domains, we will use also use the naming convention for these rhombohedral domains. Specifically, we will designate these domains as $C, A_1, A_2, A_3$. Each of these domains have spontaneous polarization directions of [111] , [$\bar{1}$11] , [1$\bar{1}$1] , [$\bar{1}\bar{1}$1] correspondingly. The notation $c$ stands for the fact that spontaneous polarization is parallel to the basis vector $\boldsymbol{c}_H$ of the hexagonal cell setting. The notation $A_i$ stands for the fact that polarization direction is parallel to the plane of vectors $\boldsymbol{c}_H$ and $\boldsymbol{a}_{mi}$.

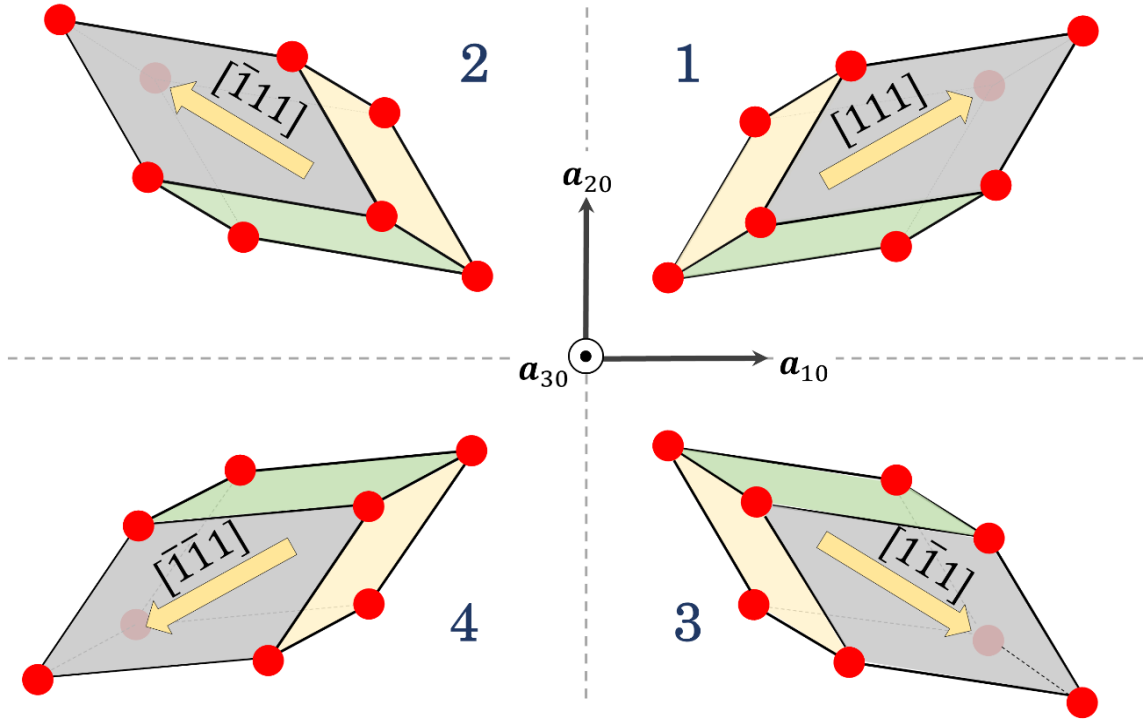

**Figure 4** The definition and numbering of four rhombohedral domains variants. The polarization directions are given relative to the basis vectors $\boldsymbol{a}_{im}$. The basis vectors of the paraelastic phase $\boldsymbol{a}_{i0}$ are shown in the figure.

**Table 3** The definitions of ferroelastic trigonal domains (organized in the same way as Table 1).

| Domain variant (name) | 1 (C) | 2 ($A_1$) | 3 ($A_2$) | 4 ($A_3$) |
|---|---|---|---|---|
| $[G]_m$ | $a^2\begin{pmatrix}1 & \eta & \eta \\ \eta & 1 & \eta \\ \eta & \eta & 1\end{pmatrix}$ | $a^2\begin{pmatrix}1 & \bar{\eta} & \bar{\eta} \\ \bar{\eta} & 1 & \eta \\ \bar{\eta} & \eta & 1\end{pmatrix}$ | $a^2\begin{pmatrix}1 & \bar{\eta} & \eta \\ \bar{\eta} & 1 & \bar{\eta} \\ \eta & \bar{\eta} & 1\end{pmatrix}$ | $a^2\begin{pmatrix}1 & \eta & \bar{\eta} \\ \eta & 1 & \bar{\eta} \\ \bar{\eta} & \bar{\eta} & 1\end{pmatrix}$ |
| $[P]_m$ | [1 1 1] | [$\bar{1}$ 1 1] | [1 $\bar{1}$ 1] | [$\bar{1}$ $\bar{1}$ 1] |

Six pairs may be formed between four ferroelastic domain variants. We will demonstrate the connection between domains $1(C)$ and $2(A_1)$:

$$[\Delta G] = [G]_2 - [G]_1 = -2a^2\eta \begin{pmatrix} 0 & 1 & 1 \\ 1 & 0 & 0 \\ 1 & 0 & 0 \end{pmatrix} \tag{43}$$

The eigenvalues and eigenvectors of (43) are such that

$$\lambda_1 = -2a^2\eta\sqrt{2}, \; \lambda_2 = 0, \; \lambda_3 = -\lambda_1$$

$$[V] = \frac{1}{2}\begin{pmatrix} -\sqrt{2} & 0 & +\sqrt{2} \\ -1 & -\sqrt{2} & -1 \\ -1 & +\sqrt{2} & -1 \end{pmatrix} \tag{44}$$

The equation (20) $(V_{i1} \pm V_{i3})x_i = 0$ takes the form

$$\begin{cases} x_2 + x_3 = 0 & \text{for the case of } (V_{i1} + V_{i3})x_i = 0 \\ x_1 = 0 & \text{for the case of } (V_{i1} - V_{i3})x_i = 0 \end{cases} \tag{45}$$

The first part of the equation (45) corresponds to the (011) plane, while the second part corresponds to (100) one. This is well-known result (Fousek & Janovec, 1969), indicating that rhombohedral domains may pair along the families of domains walls with Miller indices {011} and {100}.

**For the case of (011) domain walls,** the transformation matrix $[W] = [V][Z]$ will have to be introduced with $[Z] = \begin{bmatrix} 1 & 0 & 0 \\ 0 & 1 & 0 \\ -1 & 0 & 1 \end{bmatrix}$:

$$[W] = \frac{1}{2}\begin{pmatrix} -2\sqrt{2} & 0 & \sqrt{2} \\ 0 & -\sqrt{2} & -1 \\ 0 & \sqrt{2} & -1 \end{pmatrix} \tag{46}$$

To formulate the system of equations (24) we will first get $[G^{(W)}]_{1,2} = [W]^T [G]_{1,2} [W]$:

$$[G^{(W)}]_1 = a^2 \begin{pmatrix} 2 & 0 & \sqrt{2}\eta - 1 \\ 0 & 1 - \eta & 0 \\ \sqrt{2}\eta - 1 & 0 & \left(\frac{1}{2} - \sqrt{2}\right)\eta + 1 \end{pmatrix},$$

$$[G^{(W)}]_2 = a^2 \begin{pmatrix} 2 & 0 & -\sqrt{2}\eta - 1 \\ 0 & 1 - \eta & 0 \\ -\sqrt{2}\eta - 1 & 0 & \left(\frac{1}{2} + \sqrt{2}\right)\eta + 1 \end{pmatrix}$$

(47)

According to (22), $y_3 = 1$ and applying the system of equations (24) we get

$$\begin{cases} y_1 = -\sqrt{2}\eta \\ y_2 = 0 \end{cases}, \quad [S_w] = \begin{pmatrix} 1 & 0 & -\sqrt{2}\eta \\ 0 & 1 & 0 \\ 0 & 0 & 1 \end{pmatrix}, \tag{48}$$

According to (26) and (46) we get for the matrix $[S] = [W][S_w][W]^{-1}$:

$$[S] = \begin{pmatrix} 1 & -2\eta & -2\eta \\ 0 & 1 & 0 \\ 0 & 0 & 1 \end{pmatrix}, \quad [S^*] = \begin{pmatrix} 1 & 0 & 0 \\ 2\eta & 1 & 0 \\ 2\eta & 0 & 1 \end{pmatrix} \tag{49}$$

and

$$[S^*] - [I] = 2\eta \begin{pmatrix} 0 & 0 & 0 \\ 1 & 0 & 0 \\ 1 & 0 & 0 \end{pmatrix} \tag{50}$$

Which yields the following expression for the separation between the Bragg peaks

$$\begin{pmatrix} \Delta H \\ \Delta K \\ \Delta L \end{pmatrix} = 2\eta H \begin{pmatrix} 0 \\ 1 \\ 1 \end{pmatrix} \tag{51}$$

**For the case of $(100)$ domain wall,** the transformation matrix $[W] = [V][Z]$ will have to be introduced with $[Z] = \begin{bmatrix} 1 & 0 & 0 \\ 0 & 1 & 0 \\ 1 & 0 & 1 \end{bmatrix}$:

$$[W] = \frac{1}{2}\begin{pmatrix} 0 & 0 & \sqrt{2} \\ -2 & -\sqrt{2} & -1 \\ -2 & \sqrt{2} & -1 \end{pmatrix} \tag{52}$$

so that

$$[G^{(W)}]_1 = a^2 \begin{pmatrix} 2(\eta+1) & 0 & (1-\sqrt{2})\eta+1 \\ 0 & 1-\eta & 0 \\ (1-\sqrt{2})\eta+1 & 0 & \left(\frac{1}{2}-\sqrt{2}\right)\eta+1 \end{pmatrix}, \tag{53}$$

$$[G^{(W)}]_2 = a^2 \begin{pmatrix} 2(\eta+1) & 0 & (1+\sqrt{2})\eta+1 \\ 0 & 1-\eta & 0 \\ (1+\sqrt{2})\eta+1 & 0 & \left(\frac{1}{2}+\sqrt{2}\right)\eta+1 \end{pmatrix}$$

According to (22), $y_3 = 1$ and applying equation (24) we get

$$\begin{cases} y_1 = \frac{\sqrt{2}\eta}{\eta+1} \\ y_2 = 0 \end{cases}, \quad [S_w] = \begin{pmatrix} 1 & 0 & \frac{\sqrt{2}\eta}{\eta+1} \\ 0 & 1 & 0 \\ 0 & 0 & 1 \end{pmatrix}, \tag{54}$$

and according to the equation (26) and (46) we can get for the matrix $[S]$:

$$[S] = \begin{pmatrix} 1 & 0 & 0 \\ -\xi & 1 & 0 \\ -\xi & 0 & 1 \end{pmatrix}, \quad [S^*] = \begin{pmatrix} 1 & \xi & \xi \\ 0 & 1 & 0 \\ 0 & 0 & 1 \end{pmatrix} \tag{55}$$

Here the following notation was introduced

$$\xi = \frac{2\eta}{\eta + 1} \tag{56}$$

Accordingly

$$[S^*] - [I] = \xi \begin{pmatrix} 0 & 1 & 1 \\ 0 & 0 & 0 \\ 0 & 0 & 0 \end{pmatrix} \tag{57}$$

Which finally gives the following expression for the separation between peaks

$$\begin{pmatrix} \Delta H \\ \Delta K \\ \Delta L \end{pmatrix} = \xi(K + L) \begin{pmatrix} 1 \\ 0 \\ 0 \end{pmatrix} \tag{58}$$

Identical analysis can be implemented to the domain wall and other pairs of domains. Table 4 summarizes the results: it includes all the possible domains walls between possible domain variants $m$ and $n$ and corresponding separation of Bragg reflections in the reciprocal coordinate system of the domain $m$. As can be seen, the separation vector is always perpendicular to the domain wall.

**Table 4** The same as Table 2 but for the rhombohedral domains

| $m$ | $n$ | Plane | $[S]$ | $[S^*] - [I]$ | $\Delta \mathbf{B}$ |
|---|---|---|---|---|---|
| 1 | 2 | (011) | $\begin{pmatrix} 1 & -2\eta & -2\eta \\ 0 & 1 & 0 \\ 0 & 0 & 1 \end{pmatrix}$ | $2\eta \begin{pmatrix} 0 & 0 & 0 \\ 1 & 0 & 0 \\ 1 & 0 & 0 \end{pmatrix}$ | $2\eta H \begin{pmatrix} 0 \\ 1 \\ 1 \end{pmatrix}$ |
| 1 | 2 | (100) | $\begin{pmatrix} 1 & 0 & 0 \\ -\xi & 1 & 0 \\ -\xi & 0 & 1 \end{pmatrix}$ | $\xi \begin{pmatrix} 0 & 1 & 1 \\ 0 & 0 & 0 \\ 0 & 0 & 0 \end{pmatrix}$ | $\xi(K + L) \begin{pmatrix} 1 \\ 0 \\ 0 \end{pmatrix}$ |
| 1 | 3 | (101) | $\begin{pmatrix} 1 & 0 & 0 \\ -2\eta & 1 & -2\eta \\ 0 & 0 & 1 \end{pmatrix}$ | $2\eta \begin{pmatrix} 0 & 1 & 0 \\ 0 & 0 & 0 \\ 0 & 1 & 0 \end{pmatrix}$ | $2\eta K \begin{pmatrix} 1 \\ 0 \\ 1 \end{pmatrix}$ |
| 1 | 3 | (010) | $\begin{pmatrix} 1 & -\xi & 0 \\ 0 & 1 & 0 \\ 0 & -\xi & 1 \end{pmatrix}$ | $\xi \begin{pmatrix} 0 & 0 & 0 \\ 1 & 0 & 1 \\ 0 & 0 & 0 \end{pmatrix}$ | $\xi(H + L) \begin{pmatrix} 0 \\ 1 \\ 0 \end{pmatrix}$ |
| 1 | 4 | (110) | $\begin{pmatrix} 1 & -2\eta & -2\eta \\ 0 & 1 & 0 \\ 0 & 0 & 1 \end{pmatrix}$ | $2\eta \begin{pmatrix} 0 & 0 & 1 \\ 0 & 0 & 1 \\ 0 & 0 & 0 \end{pmatrix}$ | $2\eta L \begin{pmatrix} 1 \\ 1 \\ 0 \end{pmatrix}$ |
| 1 | 4 | (001) | $\begin{pmatrix} 1 & 0 & -\xi \\ 0 & 1 & -\xi \\ 0 & 0 & 1 \end{pmatrix}$ | $\xi \begin{pmatrix} 0 & 0 & 0 \\ 0 & 0 & 0 \\ 1 & 1 & 0 \end{pmatrix}$ | $\xi(H + K) \begin{pmatrix} 0 \\ 0 \\ 1 \end{pmatrix}$ |
| 2 | 3 | ($\bar{1}$10) | $\begin{pmatrix} 1 & 0 & 0 \\ 0 & 1 & 0 \\ 2\eta & -2\eta & 1 \end{pmatrix}$ | $2\eta \begin{pmatrix} 0 & 0 & \bar{1} \\ 0 & 0 & 1 \\ 0 & 0 & 0 \end{pmatrix}$ | $2\eta L \begin{pmatrix} \bar{1} \\ 1 \\ 0 \end{pmatrix}$ |
| 2 | 3 | (001) | $\begin{pmatrix} 1 & 0 & \xi \\ 0 & 1 & -\xi \\ 0 & 0 & 1 \end{pmatrix}$ | $\xi \begin{pmatrix} 0 & 0 & 0 \\ 0 & 0 & 0 \\ \bar{1} & 1 & 0 \end{pmatrix}$ | $\xi(K - H) \begin{pmatrix} 0 \\ 0 \\ 1 \end{pmatrix}$ |
| 2 | 4 | ($\bar{1}$01) | $\begin{pmatrix} 1 & 0 & 0 \\ 2\eta & 1 & -2\eta \\ 0 & 0 & 1 \end{pmatrix}$ | $2\eta \begin{pmatrix} 0 & \bar{1} & 0 \\ 0 & 0 & 0 \\ 0 & 1 & 0 \end{pmatrix}$ | $2\eta K \begin{pmatrix} \bar{1} \\ 0 \\ 1 \end{pmatrix}$ |

| | | | | | | |
|---|---|---|---|---|---|---|
| 2 | 4 | (010) | $\begin{pmatrix} 1 & \xi & 0 \\ 0 & 1 & 0 \\ 0 & -\xi & 1 \end{pmatrix}$ | $\xi \begin{pmatrix} 0 & 0 & 0 \\ \bar{1} & 0 & 1 \\ 0 & 0 & 0 \end{pmatrix}$ | $\xi(L-H)$ | $\begin{pmatrix} 0 \\ 1 \\ 0 \end{pmatrix}$ |
| 3 | 4 | (0$\bar{1}$1) | $\begin{pmatrix} 1 & 2\eta & -2\eta \\ 0 & 1 & 0 \\ 0 & 0 & 1 \end{pmatrix}$ | $2\eta \begin{pmatrix} 0 & 0 & 0 \\ \bar{1} & 0 & 0 \\ 1 & 0 & 0 \end{pmatrix}$ | $2\eta H$ | $\begin{pmatrix} 0 \\ \bar{1} \\ 1 \end{pmatrix}$ |
| 3 | 4 | (100) | $\begin{pmatrix} 1 & 0 & 0 \\ \xi & 1 & 0 \\ -\xi & 0 & 1 \end{pmatrix}$ | $\xi \begin{pmatrix} 0 & \bar{1} & 1 \\ 0 & 0 & 0 \\ 0 & 0 & 0 \end{pmatrix}$ | $\xi(L-K)$ | $\begin{pmatrix} 1 \\ 0 \\ 0 \end{pmatrix}$ |

## 7. Experimental method: three-dimensional high-resolution reciprocal space mapping

The experimental details of high-resolution reciprocal space mapping were explained elsewhere (Gorfman *et al.*, 2020; Zhang *et al.*, 2018). The technique uses parallel and monochromatic X-ray beam alongside with high-resolution pixel area detector. The goal of the experiment is to reconstruct the fine details of the diffraction intensity distribution around specific Bragg peaks. It allows measuring the separations of nearly overlapping Bragg peaks components (each corresponding to a separate domain). The intensity distribution in the reciprocal space is reconstructed by rotating the crystal around one of the diffractometer axis (e.g. $\omega$) and converting three coordinates $X_d\, Y_d\, \omega$ ($X_d Y_d$ are the coordinates of the detector pixels) to the coordinates of the scattering vector **B** relative to the chosen Cartesian $(B_x, B_y, B_z)$ or reciprocal crystallographic $(B_H, B_K, B_L)$ coordinate system. Such experiments are facilitated by the recent progress in synchrotron-based and home-laboratory X-ray sources, availability of pixel area detectors, beam conditioning systems and big-data exchange protocols (see e.g. (Dyadkin *et al.*, 2016; Girard *et al.*, 2019; Gorfman *et al.*, 2021)).

## 8. Recognition of coherent twin relationship in tetragonal BaTiO$_3$ crystal.

The chapter illustrates the recognition of coherent twin relationship in twinned BaTiO$_3$ crystal. We performed high-resolution reciprocal space mapping measurements at the dedicated home-laboratory X-ray diffractometer in Tel Aviv University (Gorfman *et al.*, 2021). After the determination of the average orientation matrix using CrysAlisPro software (the averaging is performed over all the domains present in the X-ray beam), we collected high-resolution reciprocal space maps of the diffraction intensity distribution around 102, 002, 222, and 103 reflections. The data was then represented in the form of three-dimensional diffraction intensity tables $I(B_x, B_y, B_z)$. Here $B_x, B_y, B_z$ refer to the Cartesian coordinate system such that **X**-axis is nearly parallel to the scattering vector, so that $B_x$ is nearly equal to the scattering vector length. Figure 5 shows $B_x, B_y$ projections $I_z(B_x, B_y) = \int I(B_x, B_y, B_z)dB_z$. Such projections visually demonstrate the separation of sub-peaks along **X**-axis (nearly equivalent to the separation along $2\theta$ axis). This splitting of the peaks along **X**-axis can be used to determine tetragonality or (in more general case) the deviation of the lattice parameters from that of the cubic system.

The initial assignment of the sub-peaks to domains can be solely based on the analysis of the scattering vector lengths. The procedure was described in (Gorfman *et al.*, 2020). It includes measuring the lengths of the scattering vector of all the observed sub-peak using the equation $|\boldsymbol{B}_{obs}| = \sqrt{B_x^2 + B_y^2 + B_z^2}$. These are matched with the calculated lengths of the reciprocal lattice vectors $|\boldsymbol{B}_{calc}| = \sqrt{G^*_{mij}H_jH_j}$ ($H_1 = H, H_2 = K, H_3 = L$ are the indices of the reflection). The lattice parameters are adjusted to achieve the best matching between the sets of $|\boldsymbol{B}_{obs}|$ and $|\boldsymbol{B}_{calc}|$. Using combined analysis of the scattering vectors lengths in four reciprocal space maps, we obtained tetragonal lattice parameters $a = 3.962(1)$, $c = 4.005(2)$ Å. According to (38) $\tau = 0.011(1)$. The white lines in the $I_z(B_x, B_y)$ projections in the Figure 5 follow the equation $\sqrt{B_x^2 + B_y^2 + B_z^2} = |\boldsymbol{B}_{calc}|$, where $B_z$ correspond to the corresponding coordinate of the best matching sub-peak.

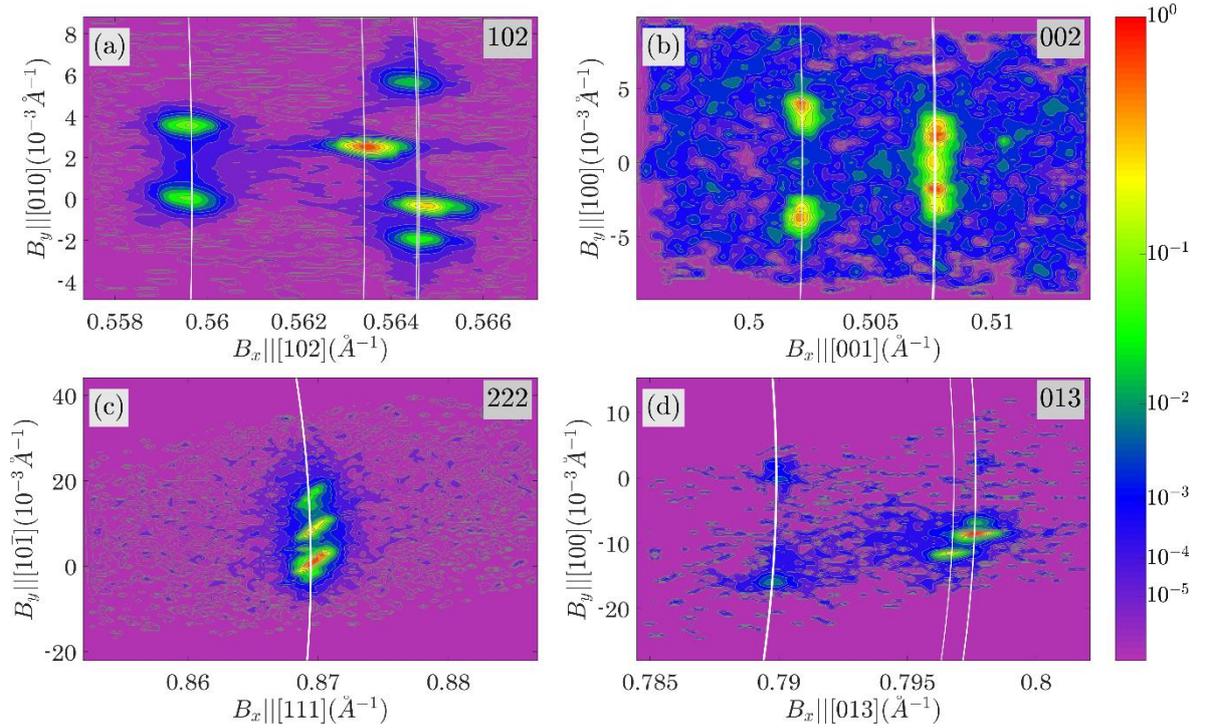

**Figure 5** $I_z(B_x, B_y)$ projections of the reciprocal space maps of 102, 002, 222 and 013 reflections from BaTiO$_3$ crystal containing ferroelastic domain of tetragonal symmetry. The white lines correspond to the equation $\sqrt{B_x^2 + B_y^2 + B_z^2} = |\boldsymbol{B}_{calc}|$ (here $|\boldsymbol{B}_{calc}|$ was calculated using tetragonal lattice parameters $a = 3.962(1)$, $c = 4.005(2)$ Å).

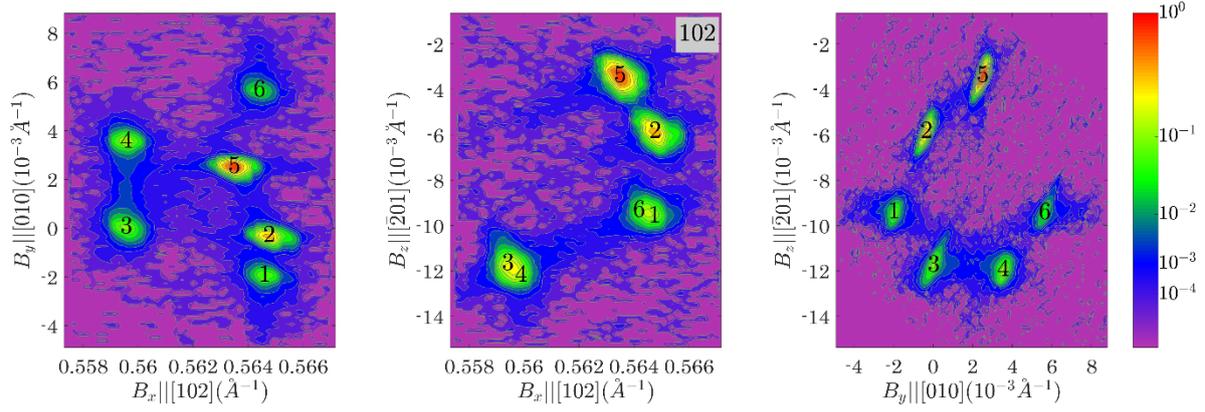

**Figure 6** $I_z(B_x B_y)$, $I_y(B_x B_z)$ and $I_x(B_y B_z)$ projections of three-dimensional diffraction intensity distribution $I(B_x, B_y, B_z)$ around 1 0 2 family of Bragg peaks of BaTiO$_3$. Six sub-peaks are located and numbered in the maps.

The example below demonstrates the assignment of peaks to domains and identification of coherent twin relationship in the corresponding reciprocal space maps. Figure 6 presents the corresponding $I_z(B_x B_y), I_y(B_x B_z)$ and $I_x(B_y, B_z)$ projections. The same figure shows the marked positions of the peaks (according to the procedure in (Gorfman *et al.*, 2020)). Table 5 summarizes the results of this marking. For each marked peak, it includes the observed and calculated length of the reciprocal lattice vector (e.g. $|\boldsymbol{B}_{obs}| = \sqrt{B_x^2 + B_y^2 + B_z^2}$ and $|\boldsymbol{B}_{calc}| = \sqrt{G_{mij} H_i H_j}$) as well as one possible assignment of peaks to the domains (the domain number(s) $m$ for which the best matching between $|\boldsymbol{B}_{obs}|$ and $|\boldsymbol{B}_{calc}|$ is achieved). Finally, the last three rows of the Table 5 illustrate the reciprocal lattice coordinates of the peaks with respect to the mass centre. Similar analysis of three other reciprocal space maps are presented in the Supporting materials.

**Table 5** The summary of the individual sub-peaks marked in the 012 reciprocal space map. The top row gives the peaks numbers, the second row gives the corresponding length of the reciprocal space vector ($|\boldsymbol{B}_{obs}| = \sqrt{B_x^2 + B_y^2 + B_z^2}$), the third row gives the best matching calculated length of the reciprocal lattice vector ($|\boldsymbol{B}_{calc}| = \sqrt{G^*_{mij} H_i H_j}$, the forth row gives the domain number(s) $m$ for which this matching is achieved. Three bottom rows give the reciprocal lattice coordinates ($\Delta \boldsymbol{B} = \Delta B_1 \boldsymbol{a}_1^* + \Delta B_2 \boldsymbol{a}_2^* + \Delta B_3 \boldsymbol{a}_3^*$) of all the peaks with respect to the peak centre of gravity.

| Sub-peak number | 1 | 2 | 3 | 4 | 5 | 6 |
| --- | --- | --- | --- | --- | --- | --- |
| $|\boldsymbol{B}_{obs}|$, Å$^{-1}$ | 0.5645 | 0.5648 | 0.5595 | 0.5598 | 0.5635 | 0.5645 |
| $|\boldsymbol{B}_{calc}|$, Å$^{-1}$ | 0.5646 | 0.5646 | 0.5598 | 0.5598 | 0.5634 | 0.5646 |

| Domain assignments | 2 | 2 | 3 | 3 | 1 | 2 |
|---|---|---|---|---|---|---|
| $\Delta B_1$ ($10^{-2}$) | 1.4 | 0.2 | 1.3 | 1.4 | -0.9 | 1.4 |
| $\Delta B_2$ ($10^{-2}$) | -1.4 | -0.7 | -0.6 | 0.8 | 0.4 | 1.7 |
| $\Delta B_3$ ($10^{-2}$) | -0.1 | 0.6 | -2.3 | -2.4 | 0.6 | -0.1 |

**Table 6  Identification of coherent twin relationship using 102 families of Bragg peaks.** The first two columns show domain numbers. The third and fours column indicate the expected separation between the peaks in the analytical and numerical form correspondingly. The last column shows the best matching (when such matching is found) separation between the Bragg peaks and their numbers according to the Table 5.

| $m$ | $n$ | $\Delta \boldsymbol{B}$ (equation) | $\Delta \boldsymbol{B}$ (calculated), $10^{-2}$ | $\Delta \boldsymbol{B}$ (measured) $10^{-2}$ | Sub peaks pair (According the numbering in Figure 6) | |
|---|---|---|---|---|---|---|
| $1(a)$ | $2(b)$ | $\tau(H+K)\begin{pmatrix}1\\\bar{1}\\0\end{pmatrix}$ | $1.1\begin{pmatrix}1\\\bar{1}\\0\end{pmatrix}$ | $\begin{pmatrix}+1.1\\-1.1\\0\end{pmatrix}$ | 5 | 2 |
| $2(b)$ | $3(c)$ | $\tau(K+L)\begin{pmatrix}0\\1\\\bar{1}\end{pmatrix}$ | $2.2\begin{pmatrix}0\\1\\\bar{1}\end{pmatrix}$ | $\begin{pmatrix}+0.0\\+2.2\\-2.3\end{pmatrix}$ | 1 | 4 |
| $2(b)$ | $3(c)$ | $\tau(K-L)\begin{pmatrix}0\\1\\1\end{pmatrix}$ | $-2.2\begin{pmatrix}0\\1\\1\end{pmatrix}$ | $\begin{pmatrix}-0.1\\-2.3\\-2.2\end{pmatrix}$ | 6 | 3 |

Figure 7 illustrates the result of the identification of the coherent twin relationship. It shows the same projections of three-dimensional intensity distribution (as in the Figure 6) with the relevant connections between the sub-peaks. The sub-peaks are connected with each other if the separation between them (the fifth column of the Table 6) matches one of the theoretically predicted (the fourth columns of the Table 6).

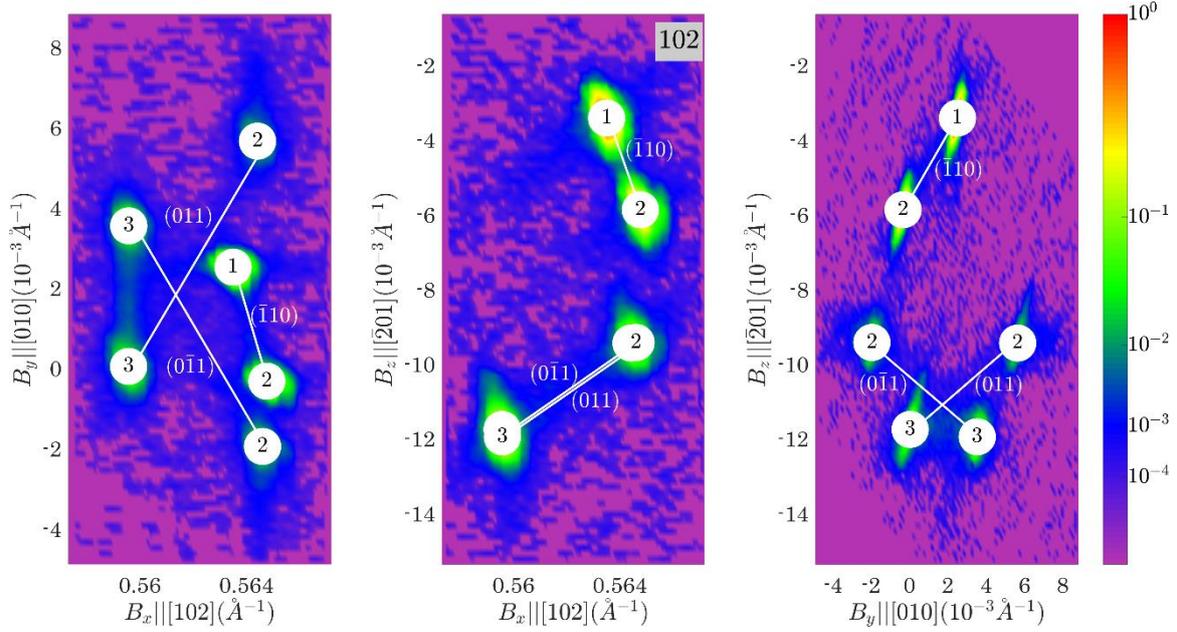

**Figure 7** The same as in the Figure 6, but after the assignment of peaks to the domains (definition and numbering of domains is presented in the Figure 6 and Table 5. The solid lines connect the peaks pairs, which correspond to the matched domains. The Miller indices of the matching plane are indicated in the bracket.

According to the Table 6 and Figure 7, all three tetragonal domains are present in the relevant volume of the crystal (exposed by the X-ray beam during the collection of this reciprocal space map). The following coherent twin relationship among them can be identified:

- $1(a)$ and $2(b)$ domains, connected to each other via $(1\bar{1}0)$ domain wall.
- $2(b)$ and $3(c)$ domains, connected to each other via $(0\bar{1}1)$ domain wall.
- $2(b)$ and $3(c)$ domains, connected to each other via $(011)$ domain wall.

## 9. Recognition of ferroelastic domains in PbZr$_{0.75}$Ti$_{0.25}$O$_3$ rhombohedral single crystal

This paragraph illustrates the recognition of coherent twin relationship in the high-resolution X-ray diffraction patterns of twinned PbZr$_{0.75}$Ti$_{0.25}$O$_3$ crystal. The data was collected exactly as for the case of BaTiO$_3$ crystal (§8) and at the custom-built diffractometer in Tel Aviv University. Figure 8 (organized as Figure 5) shows $I_z(B_x, B_y)$ projections of diffraction intensity distribution. Similar to the case of BaTiO$_3$ the separation of the peaks along the $X$-axis can be used to determine the rhombohedral distortion. The rhombohedral lattice parameters a = 4.115(1) Å, $\gamma = 89.686°$ were obtained. Accordingly $\eta = 0.0055$ and $\xi = 0.0109$.

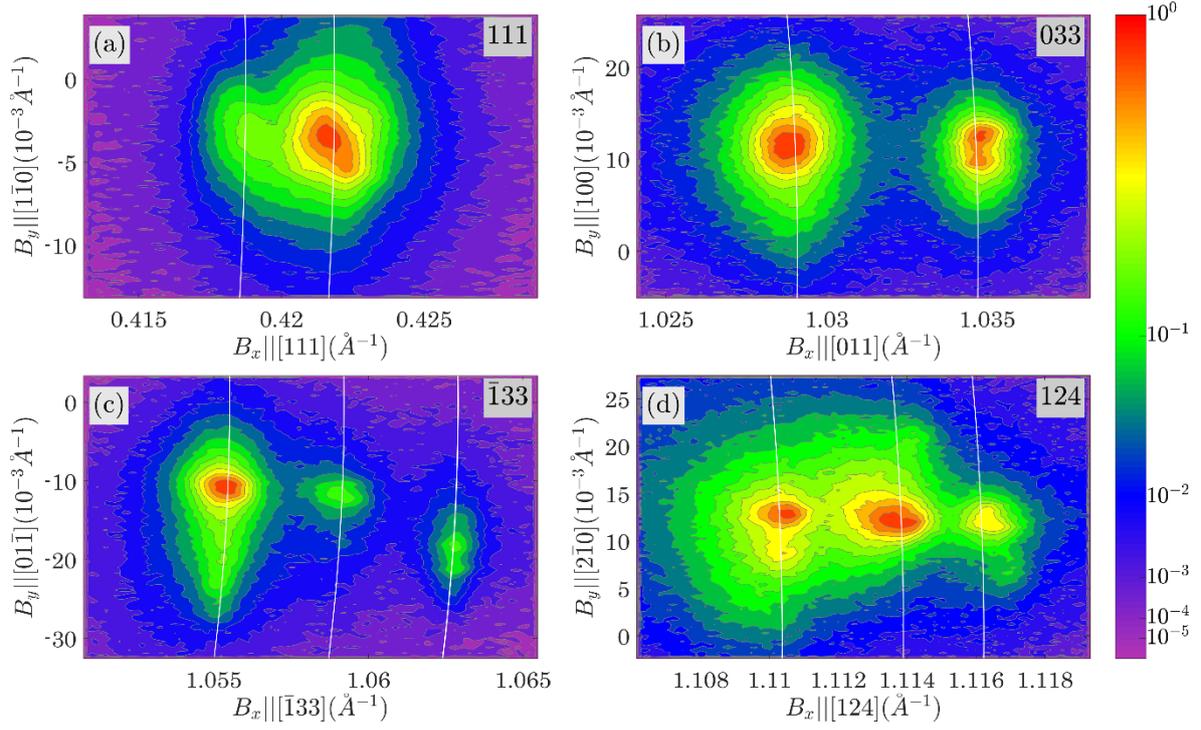

**Figure 8** The same as Figure 5 but for the reciprocal space maps of 111, 033, $\bar{1}33$ and 124 reflections from twinned PbZr$_{0.75}$Ti$_{0.25}$O$_3$ crystal containing domains of rhombohedral symmetry. The white lines correspond to the reciprocal lattice vectors lengths, calculated using rhombohedral lattice parameters $a = b = c = 4.115(1)$ Å, $\alpha = \beta = \gamma = 89.686°$.

The example below demonstrates the assignment of the peaks in the reciprocal space maps of 124 reflection. Figure 9 and Figure 10 are organized as Figure 6 and Figure 7.

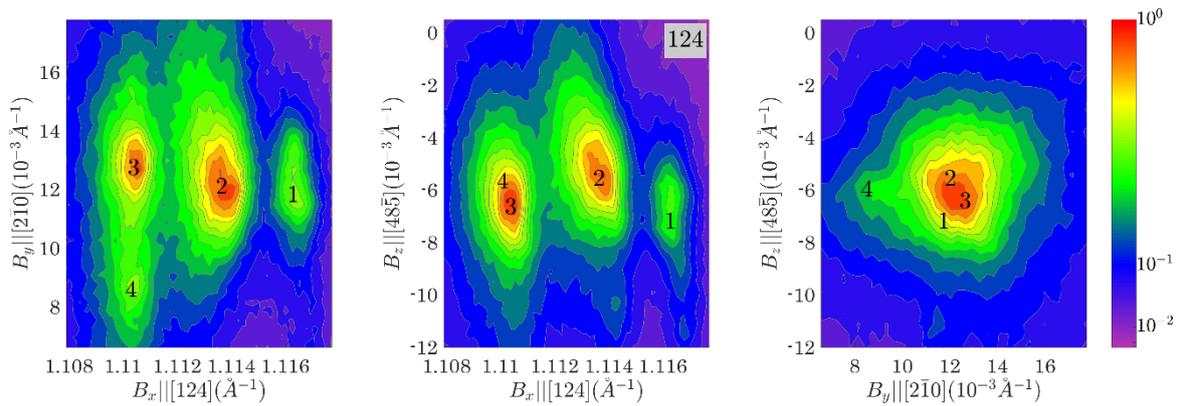

**Figure 9** $I_z(B_x B_y)$, $I_y(B_x B_z)$ and $I_x(B_y B_z)$ projections of three-dimensional diffraction intensity distribution $I(B_x, B_y, B_z)$ around 124 family of Bragg peaks of PZT. Four sub-peaks are located and numbered in the maps.

**Table 7**  The coordinates of individual sub peaks (in the reciprocal lattice units). The error bar of 0.001 is assumed for each number in the table.

| Sub-peak number | 1 | 2 | 3 | 4 |
|---|---|---|---|---|
| $|\boldsymbol{B}_{obs}|$, Å$^{-1}$ | 1.1162 | 1.1138 | 1.1105 | 1.1102 |
| $|\boldsymbol{B}_{calc}|$, Å$^{-1}$ | 1.1162 | 1.1139 | 1.1104 | 1.1104 |
| Domain assignments | 3 | 2 | 1 | 1 |
| $\Delta B_1$ ($10^{-2}$) | -0.5 | -0.4 | -0.6 | -2.1 |
| $\Delta B_2$ ($10^{-2}$) | 1.7 | 1.7 | 0.7 | 1.6 |
| $\Delta B_3$ ($10^{-2}$) | 1.6 | 0.4 | -0.6 | -0.8 |

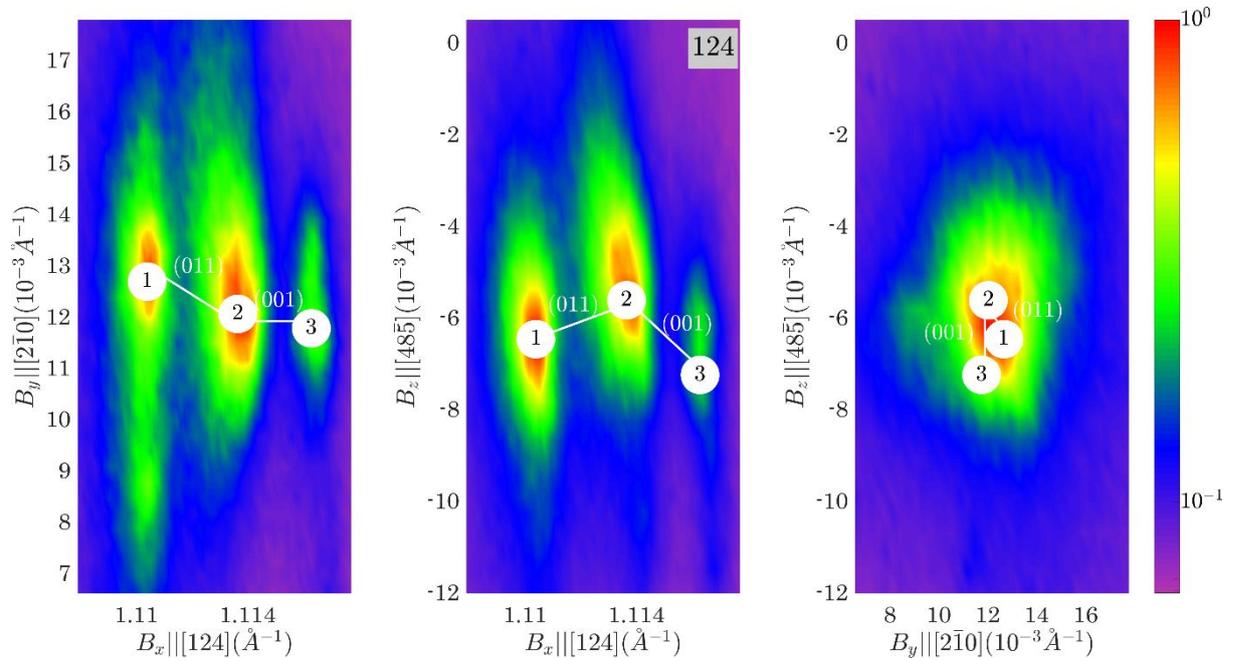

**Figure 10**    The same as in the Figure 7 but for the 124 reciprocal space map from multi-domain PbZr$_{0.75}$Ti$_{0.25}$O$_3$ crystal. The solid lines connect the peaks diffracted from the matched domains. The Miller indices of the matching plane are indicated in the brackets.

**Table 8  Identification of coherent twin relationship in the 124 reciprocal space maps of rhombohedral PbZr$_{0.75}$Ti$_{0.25}$O$_3$ crystal.** The first two columns show domain numbers (and names). The third and fours column indicate the expected separation between the peaks in the analytical and numerical form correspondingly. The last column shows the best matching separation between the Bragg peaks and their numbers according to the Table 5.

| $m$ | $n$ | $\Delta \boldsymbol{B}$ (equation) | $\Delta \boldsymbol{B}$ (calculated), $10^{-2}$ | $\Delta \boldsymbol{B}$ (measured) $10^{-2}$ | Sub peaks pairs (According the numbering in Figure 9) | |
| --- | --- | --- | --- | --- | --- | --- |
| 1(C) | 2(A$_1$) | $2\eta H \begin{pmatrix} 0 \\ 1 \\ 1 \end{pmatrix}$ | $1.1 \begin{pmatrix} 0 \\ 1 \\ 1 \end{pmatrix}$ | $\begin{pmatrix} +0.2 \\ +1.0 \\ +1.0 \end{pmatrix}$ | 3 | 2 |
| 2(A$_1$) | 3(A$_2$) | $\xi(K-H) \begin{pmatrix} 0 \\ 0 \\ 1 \end{pmatrix}$ | $1.1 \begin{pmatrix} 0 \\ 0 \\ 1 \end{pmatrix}$ | $\begin{pmatrix} -0.1 \\ +0.0 \\ +1.2 \end{pmatrix}$ | 2 | 1 |

According to the Table 8 and Figure 10, all three tetragonal domains are present in the relevant volume of the crystal. In addition, the following coherent twin relationship can be identified:

- $1(C)$ and $2(A_1)$ domains, connected via (011) domain wall.
- $2(A_1)$ and $3(A_2)$ domains, connected via (001) domain wall.

## 10. Discussion

The presented algorithm may be useful in many cases, e.g. for the investigation of the response of a multi-domain system to an external perturbation (e.g. temperature or electric field). Considering that the integrated intensities of the peaks are proportional to the volume fraction of the corresponding domains in the beam, it is possible to describe the evolution of domain pattern quantitatively. This method was used for the estimation of extrinsic and intrinsic contributions to the electromechanical coupling in PbZr$_{1-x}$Ti$_x$O$_3$ single crystal (Gorfman *et al.*, 2020). The ability of assignment of peaks to domains allows relating the corresponding change in the domain volume fraction as a function of the domain orientation, including e.g. the direction of the spontaneous polarization vector with respect to applied perturbation (e.g. electric field).

We have demonstrated the procedure of domains recognition for the cases of crystals with tetragonal and rhombohedral domains. The same algorithm can be applied e.g. to the domains of other symmetry (e.g. monoclinic symmetry and as will be demonstrated in the upcoming publication(s)). Moreover, it may also be used for the analysis of the connections of domains of different symmetry. Formation of the habit planes between domains is possible every time when at least one eigenvector of the matrix $[\Delta G]$ is zero. The implication of this condition for the cases when pairing of different phases (e.g. rhombohedral and tetragonal) are in question will be also discussed in the forthcoming publications.

It is important that the presented technique is able to recognize domain pairs, rather than domains themselves individually. Accordingly, some of the peaks may remain unrecognized. Such cases are apparent e.g. in the 002 reciprocal space maps (see the supplemental Figure S1). Unpaired peaks may appear when e.g. the limited volume of the crystal is covered by an X-ray beam (because of strong absorption of an X-ray beam hiding some domains keeping some of the peaks unpaired). This is the reason why assignment of peaks may fail in such cases. Note that assignment may still be made attempted based on the length of the reciprocal lattice vector and the radial position of the peak in the reciprocal space. In some cases, it means that suggestions of more than one domain for a single sub-peak might be possible.

## 11. Conclusions

We developed the theoretical framework for the calculation of three-dimensional splitting of Bragg peaks, diffracted from crystal with ferroelastic domains. Specifically we extended the existing theory of domains mechanical compatibility to calculate the corresponding geometry of the reciprocal space. We have shown (analytically) that Bragg peaks always separate along the reciprocal space direction that is perpendicular to the domain wall. The analytical expression for the Bragg peak separation for the cases of the entire domain wall between domains of tetragonal and rhombohedral symmetry were obtained. The formalism is illustrated on the example of single-crystal X-ray diffraction from multi-domain $BaTiO_3$ crystal with tetragonal domains and multi-domain $PbZr_{0.75}Ti_{0.25}O_3$ crystal with rhombohedral domains. It can be useful for the analysis of the individual domains response to external perturbation (e.g. change of the temperature or external electric field).

**Acknowledgements** We acknowledge Prof Yachin Ivry (Israel Institute of Technology) for providing $BaTiO_3$ crystal used in the §8.

**Appendix A. Derivation of equation (2)**

This appendix proves the equation (2) for the relationship between the metric tensor of domain 1 and $m$, related by the twinning matrix $[T]_m$. To do this let us introduce the matrix $[U_A]_1$, describing the crystallographic coordinate system $\boldsymbol{a}_{i1}$, relative to the Cartesian coordinate system $\boldsymbol{e}_i$. The columns of this matrix represent the coordinates of the matrix $\boldsymbol{a}_{11}, \boldsymbol{a}_{21}, \boldsymbol{a}_{31}$ relative to the vectors $\boldsymbol{e}_1, \boldsymbol{e}_2, \boldsymbol{e}_3$. This means the following formal matrix equation is valid:

$$(\boldsymbol{a}_{11} \quad \boldsymbol{a}_{21} \quad \boldsymbol{a}_{31}) = (\boldsymbol{e}_1 \quad \boldsymbol{e}_2 \quad \boldsymbol{e}_3)[U_A]_1 \tag{59}$$

Here, we do not consider any orientation relationship between the paraelastic and ferroelastic phase (even if such exist) therefore the elements of $[U_A]_1$ are generally unknown. However, we just assume that domain $m$ has the same orientation relationship as domain 1 just with respect to the Cartesian coordinate system $e_{im}$ (see equation (1)). This means that the lattice of the domain $m$ can be described by the basis vectors $a'_{im}$ such that

$$(a'_{1m} \quad a'_{2m} \quad a'_{3m}) = (e_{1m} \quad e_{2m} \quad e_{3m})[U_A]_1 \tag{60}$$

Considering the definition of a twinning matrix $[T]_m$ (1) we can rewrite (60) in the form

$$(a'_{1m} \quad a'_{2m} \quad a'_{3m}) = (e_1 \quad e_2 \quad e_3)[T]_m[U_A]_1 \tag{61}$$

We will now transform the basis vectors of the domain $m$ ($a'_{im}$ to $a_{im}$) so that $a_{im}$ are nearly parallel to the vectors $a_{i1}$. Such transformation can be done using the matrix $[[T]_m]^{-1}$ (this matrix would e.g. convert the basis vectors of the Cartesian coordinate system $e'_{im}$ to $e_{im}$) so that

$$(a_{1m} \quad a_{2m} \quad a_{3m}) = (e_1 \quad e_2 \quad e_3)[T]_m[U_A]_1[[T]_m]^{-1} \tag{62}$$

Accordingly, we derive the following relationship:

$$[U_A]_m = [T]_m[U_A]_1[[T]_m]^{-1} \tag{63}$$

Finally, let us consider that the metric tensor can be calculated as $[G] = [U_A]^T[U_A]$. Considering the orthogonality condition, $[[T]_m]^{-1} = [[T]_m]^T$ we immediately arrive to the relationship (2).

# Supporting information

**S1. Recognition of peaks in three-dimensional reciprocal space maps of BaTiO$_3$.**

This supporting information demonstrates the assignments of sub-peaks to domains in the reciprocal space maps of tetragonal BaTiO$_3$ crystal. The organization of figures and tables is exactly as in §7. Figures S1, S3, S5 (identical to Figure 6) show $I_z(H_x, H_y)$, $I_y(H_x, H_z)$ and $I_x(H_y, H_z)$ projections of diffraction intensity distribution around 002, 222 and 013 families of Bragg peaks correspondingly. The individual sub-peaks are explicitly marked. Tables S1, S3 and S5 (identical to Table 5) summarize the positions of individual sub-peaks in the corresponding figures. Table S2, S4, S6 (identical to Table 6) illustrate the assignment of the peaks / peaks pairs to the matched domain pairs. These connections are shown in Figures S2, S4, and S6 (identical to Figure 7).

**S1.1. Example 1: splitting of 002 reflection / BaTiO$_3$.**

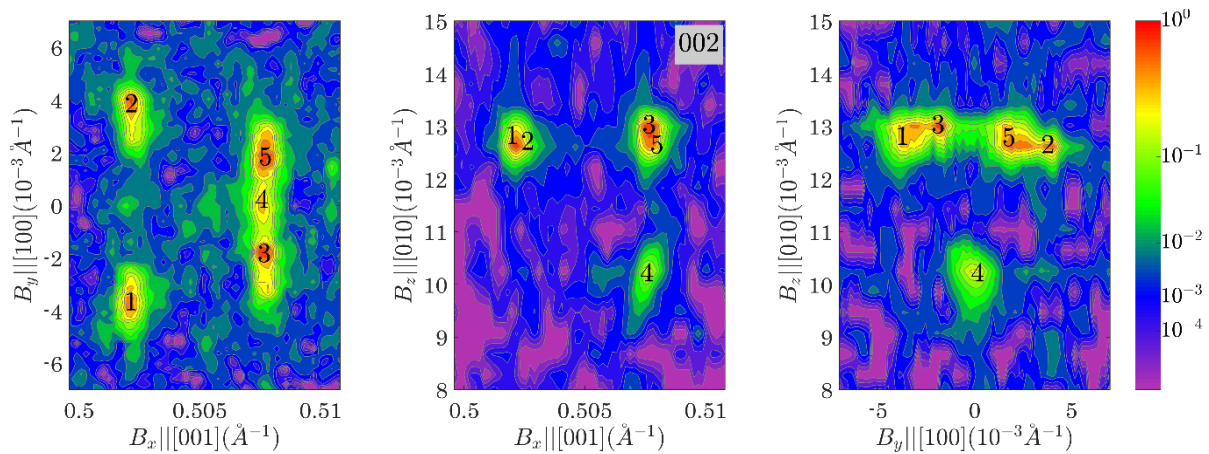

**Figure S1** The same as Figure 6 but for the case of intensity distribution around 002 reflection from multi-domain BaTiO$_3$ crystal.

**Table S1** The summary of the individually marked sub-peaks in the 002 reciprocal space map. The top row gives the peaks numbers (per their marking in Figure S1), the second row gives the corresponding length of the reciprocal space vector ($|\boldsymbol{B}_{obs}| = \sqrt{B_x^2 + B_y^2 + B_z^2}$), the third row gives the best matching calculated length of the reciprocal lattice vector ($|\boldsymbol{B}_{calc}| = \sqrt{G^*_{mij} H_i H_j}$), the forth row gives the domain number(s) $m$ for which this matching is achieved. Three bottom rows give the reciprocal lattice coordinates ($\Delta \boldsymbol{B} = \Delta B_1 \boldsymbol{a}_1^* + \Delta B_2 \boldsymbol{a}_2^* + \Delta B_3 \boldsymbol{a}_3^*$) of all the peaks relative to the peak center of gravity

| Sub-peak number | 1 | 2 | 3 | 4 | 5 |
| --- | --- | --- | --- | --- | --- |
| $\|B_{obs}\|$, Å$^{-1}$ | 0.5025 | 0.5025 | 0.5080 | 0.5075 | 0.5080 |
| $\|B_{calc}\|$, Å$^{-1}$ | 0.5025 | 0.5025 | 0.5079 | 0.5079 | 0.5079 |
| Domain assignments | 3(c) | 3(c) | 1(a), 2(b) | 1(a), 2(b) | 1(a), 2(b) |
| $\Delta B$ $\Delta B_1$ (10$^{-2}$) | -1.6 | 1.4 | -0.9 | 0.0 | 0.6 |
| $\Delta B_2$ (10$^{-2}$) | -0.3 | -0.4 | -0.2 | -1.3 | -0.3 |
| $\Delta B_3$ (10$^{-2}$) | -1.3 | -1.3 | 0.9 | 0.8 | 0.9 |

**Table S2   Identification of coherent twin relationship present among 002 families of Bragg peaks.** The first two columns show domain numbers and names (per definitions in Figure 3). The third and four's column indicate the expected separation between the peaks in the analytical and numerical form correspondingly. The fifth column shows the best matching (when reasonable match is found) separation between the Bragg peaks and their numbers according to the Table S1.

| $m$ | $n$ | $\Delta B$ (equation) | $\Delta B$ (calculated), 10$^{-2}$ | $\Delta B$ (measured) 10$^{-2}$ | Sub peaks pair (According the numbering in Figure S1) | |
|---|---|---|---|---|---|---|
| 1(a) | 3(c) | $\tau(H+L)\begin{pmatrix}1\\0\\\bar{1}\end{pmatrix}$ | $+2.2\begin{pmatrix}1\\0\\\bar{1}\end{pmatrix}$ | $\begin{pmatrix}+2.3\\-0.2\\-2.2\end{pmatrix}$ | 3 | 2 |
| 1(a) | 3(c) | $\tau(H-L)\begin{pmatrix}1\\0\\1\end{pmatrix}$ | $-2.2\begin{pmatrix}1\\0\\1\end{pmatrix}$ | $\begin{pmatrix}-2.2\\+0.0\\-2.2\end{pmatrix}$ | 5 | 1 |

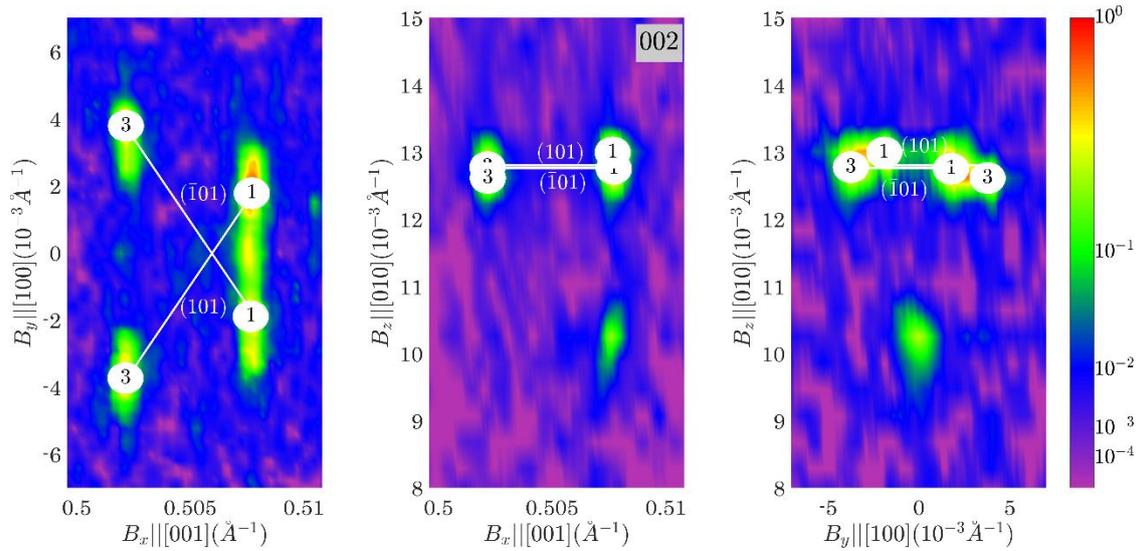

**Figure S2** The same as in the Figure 7 but for the case of 002 reflection from multi-domain BaTiO$_3$ crystal.

Accordingly, the following assignment can be made.
- The sub-peaks 3 and 2 are assigned to the tetragonal domains $1(a)$ and $3(c)$. These domains meet along $(10\bar{1})$ oriented domain wall.
- The sub-peaks 5 and 1 are assigned to the tetragonal domains $1(a)$ and $3(c)$. These domains meet along $(101)$ oriented domain wall.

Note, that no unambiguous assignment could be done for the sub-peak 4. If such assignment were made based on the length of the reciprocal lattice vector alone then this sub-peak would be associated with either domain $1(a)$ or domain $2(b)$.

### S1.2. Example 2: Splitting of 222 reflection / BaTiO$_3$

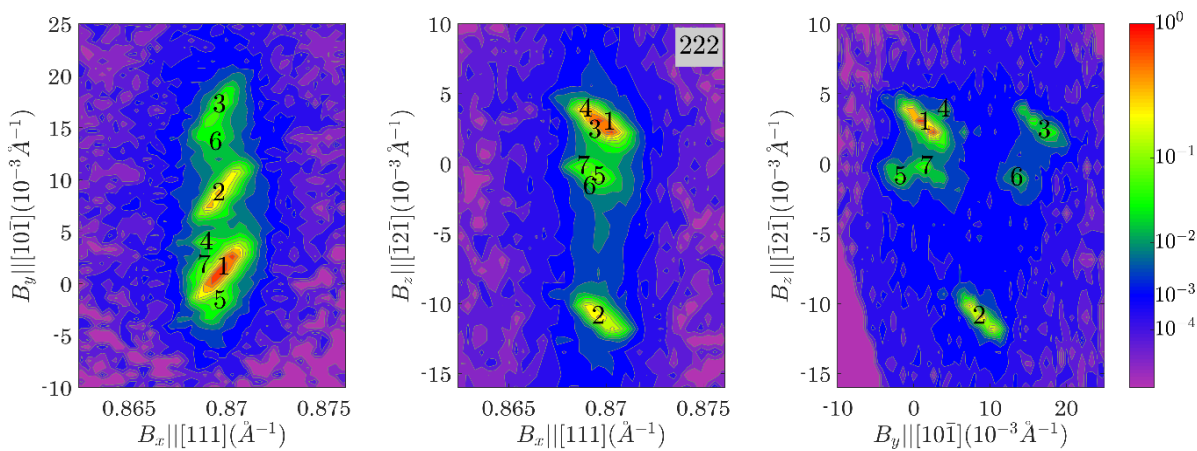

**Figure S3** The same as Figure S1 but for the case of 222 reflection from multi-domain BaTiO$_3$ crystal.

**Table S3** The same as Table S1 but for the case of 222 reflection from BaTiO$_3$ crystal.

| Sub-peak number | | 1 | 2 | 3 | 4 | 5 | 6 | 7 |
|---|---|---|---|---|---|---|---|---|
| $|B_{obs}|$, Å$^{-1}$ | | 0.8698 | 0.8695 | 0.8700 | 0.8690 | 0.8698 | 0.8695 | 0.869 |
| $|B_{calc}|$, Å$^{-1}$ | | 0.8695 | 0.8695 | 0.8695 | 0.8695 | 0.8695 | 0.8695 | 0.869 |
| Domain assignments | | all | all | all | all | all | all | All |
| $\Delta B$ | $\Delta B_1$ $(10^{-2})$ | -1.6 | 2.6 | 2.9 | -1.2 | -1.9 | 2.4 | -1.2 |

| | | | | | | | |
|---|---|---|---|---|---|---|---|
| $\Delta B_2$ $(10^{-2})$ | 1.4 | -3.2 | 1.1 | 1.4 | 0.1 | 0.0 | 0.1 |
| $\Delta B_3$ $(10^{-2})$ | 0.2 | 0.5 | -4.0 | -0.8 | 1.8 | -2.6 | 0.5 |

**Table S4** The same as Table S2 but for the case of 222 reflection from BaTiO$_3$ crystal.

| $m$ | $n$ | $\Delta B$ (equation) | $\Delta B$ (calculated), $10^{-2}$ | $\Delta B$ (measured) $10^{-2}$ | Sub peaks pair, $m$ and $n$ (According the numbering in Figure S3) | |
|---|---|---|---|---|---|---|
| $1(a)$ | $2(b)$ | $\tau(H+K)\begin{pmatrix}1\\\bar{1}\\0\end{pmatrix}$ | $+4.3\begin{pmatrix}1\\\bar{1}\\0\end{pmatrix}$ | $\begin{pmatrix}+4.2\\-4.6\\+0.3\end{pmatrix}$ | 1 | 2 |
| $1(a)$ | $3(c)$ | $\tau(H+L)\begin{pmatrix}1\\0\\\bar{1}\end{pmatrix}$ | $+4.3\begin{pmatrix}1\\0\\\bar{1}\end{pmatrix}$ | $\begin{pmatrix}+4.3\\-0.1\\-4.4\end{pmatrix}$ | 5 | 6 |
| $1(a)$ | $3(c)$ | $\tau(H+L)\begin{pmatrix}1\\0\\\bar{1}\end{pmatrix}$ | $+4.3\begin{pmatrix}1\\0\\\bar{1}\end{pmatrix}$ | $\begin{pmatrix}+4.5\\-0.3\\-4.2\end{pmatrix}$ | 1 | 3 |
| $2(b)$ | $3(c)$ | $\tau(K+L)\begin{pmatrix}0\\1\\\bar{1}\end{pmatrix}$ | $+4.3\begin{pmatrix}0\\1\\\bar{1}\end{pmatrix}$ | $\begin{pmatrix}+0.3\\+4.3\\-4.5\end{pmatrix}$ | 2 | 3 |

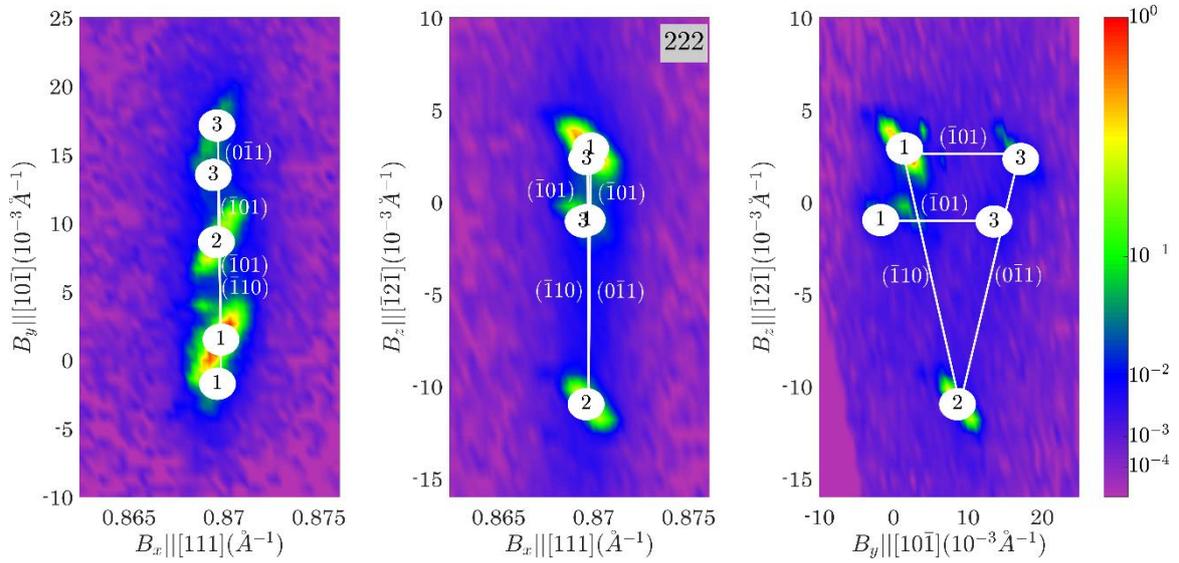

**Figure S4** The same as in the Figure S2 but for the case of 222 reflection from crystal BaTiO$_3$.

Accordingly, the following assignment can be made:

- The sub-peaks 1 and 2 are assigned to the tetragonal domains $1(a)$ and $2(b)$. These domains are matched along $(1\bar{1}0)$ oriented domain wall.

- The sub-peaks 1 and 3 are assigned to the tetragonal domains $1(a)$ and $3(c)$. These domains meet along $(10\bar{1})$ domain wall.
- The sub-peaks 5 and 6 can be assigned to the tetragonal domains $1(a)$ and $3(c)$. These domains meet along $(10\bar{1})$ domain wall.
- The sub-peaks 2 and 3 are assigned to the tetragonal domains $2(b)$ and $3(c)$. These domains meet along $(01\bar{1})$ domain wall.

### S1.3. Example 3: Splitting of 013 reflection / BaTiO$_3$

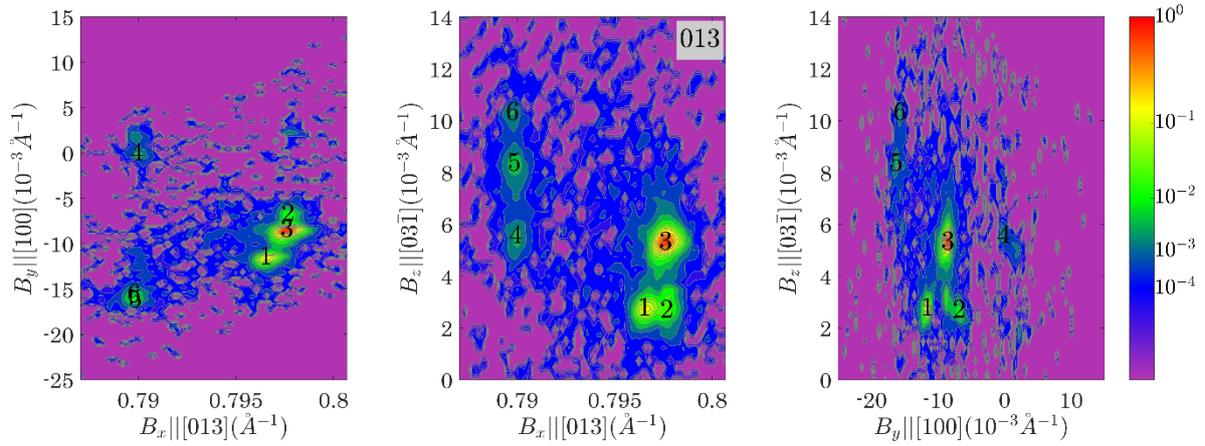

**Figure S5** The same as in the Figure S1 but for the case of 013 reflection from multi-domain BaTiO$_3$ crystal.

**Table S5** The same as Table S1 but for the case of 013 reflection from multi-domain BaTiO$_3$ crystal.

| | Sub-peak number | 1 | 2 | 3 | 4 | 5 | 6 |
|---|---|---|---|---|---|---|---|
| | $|\boldsymbol{B}_{obs}|$, Å$^{-1}$ | 0.7965 | 0.7977 | 0.7975 | 0.7900 | 0.7900 | 0.7900 |
| | $|\boldsymbol{B}_{calc}|$, Å$^{-1}$ | 0.7969 | 0.7977 | 0.7977 | 0.7900 | 0.7900 | 0.7900 |
| | Domain assignments | 2 | 1 | 1 | 3 | 3 | 3 |
| | $\Delta B_1$ ($10^{-2}$) | -1.0 | 0.9 | 0.2 | 3.6 | -2.9 | -2.7 |
| $\Delta \boldsymbol{B}$ | $\Delta B_2$ ($10^{-2}$) | -0.9 | -0.8 | 0.1 | -0.7 | 0.3 | 1.0 |
| | $\Delta B_3$ ($10^{-2}$) | 0.0 | 0.5 | 0.1 | -2.8 | -3.2 | -3.4 |

**Table S6** The same as Table S2 but for the case of 013 reflection from BaTiO$_3$ crystal.

| $m$ | $n$ | $\Delta \mathbf{B}$ (equation) | $\Delta \mathbf{B}$ (calculated), $10^{-2}$ | $\Delta \mathbf{B}$ (measured) $10^{-2}$ | Sub peaks pair, $m$ and $n$ (According the numbering in Figure S5) | |
|---|---|---|---|---|---|---|
| $1(a)$ | $2(b)$ | $\tau(H-K)\begin{pmatrix}1\\1\\0\end{pmatrix}$ | $-1.1\begin{pmatrix}1\\1\\0\end{pmatrix}$ | $\begin{pmatrix}-1.2\\-1.0\\-0.1\end{pmatrix}$ | 3 | 1 |
| $1(a)$ | $3(c)$ | $\tau(H+L)\begin{pmatrix}1\\0\\\bar{1}\end{pmatrix}$ | $3.2\begin{pmatrix}1\\0\\\bar{1}\end{pmatrix}$ | $\begin{pmatrix}+2.7\\+0.1\\-3.3\end{pmatrix}$ | 2 | 4 |
| $1(a)$ | $3(c)$ | $\tau(H-L)\begin{pmatrix}1\\0\\1\end{pmatrix}$ | $-3.2\begin{pmatrix}1\\0\\1\end{pmatrix}$ | $\begin{pmatrix}-3.1\\+0.2\\-3.3\end{pmatrix}$ | 3 | 5 |

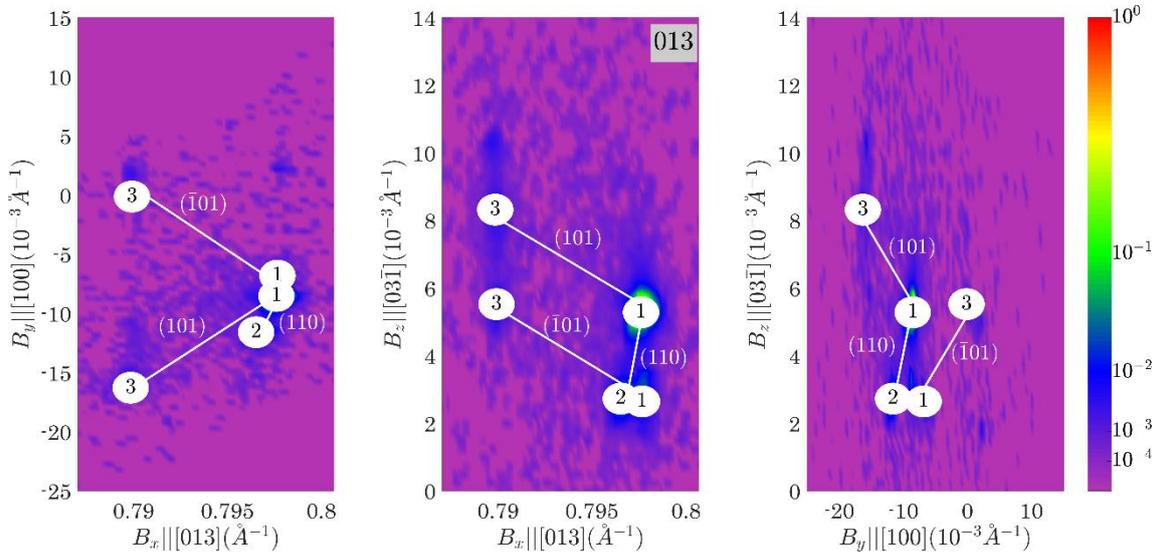

**Figure S6** The same as in the Figure S2 but for the case of 013 family of reflection.

Accordingly, the following assignment can be made:

- The sub-peaks 3 and 1 are assigned to the tetragonal domains $1(a)$ and $2(b)$. These domains meet along $(110)$ domain wall.
- The sub-peaks 2 and 4 can be assigned to the tetragonal domains $1(a)$ and $3(c)$ and these peaks are matched along $(10\bar{1})$-oriented domain wall.
- The sub-peaks 3 and 5 can be assigned to the tetragonal domains $1(a)$ and $3(c)$ and these peaks are matched along $(101)$ domain wall.

**S2. Recognition of peaks in three-dimensional reciprocal space maps of PbZr$_{0.75}$Ti$_{0.25}$O$_3$.**

This supporting information shows several examples of assignments of sub-peaks to domains in rhombohedral PZT crystal. The figures and tables are organized exactly as in §7. Specifically Figures S7, S9, S11 (identical to Figure 6) show $I_2(H_x, H_y)$, $I_2(H_x, H_z)$ and $I_2(H_y, H_z)$ projections of diffraction intensity distribution around 013, $\bar{1}33$ and 233 Bragg reflections correspondingly where individual sub-peaks are explicitly marked. Tables S7, S9 and S11 (identical to Table 5) summarize the positions of individual sub-peaks in the corresponding figures. Table S8, S10, S12 (identical to Table 6) illustrate the assignment of the peaks / peaks pairs to the matched domain pairs. These connections are illustrated in Figure S8, S10, S12.

**S2.1. Example 1: Splitting of 013 reflection / PZT**

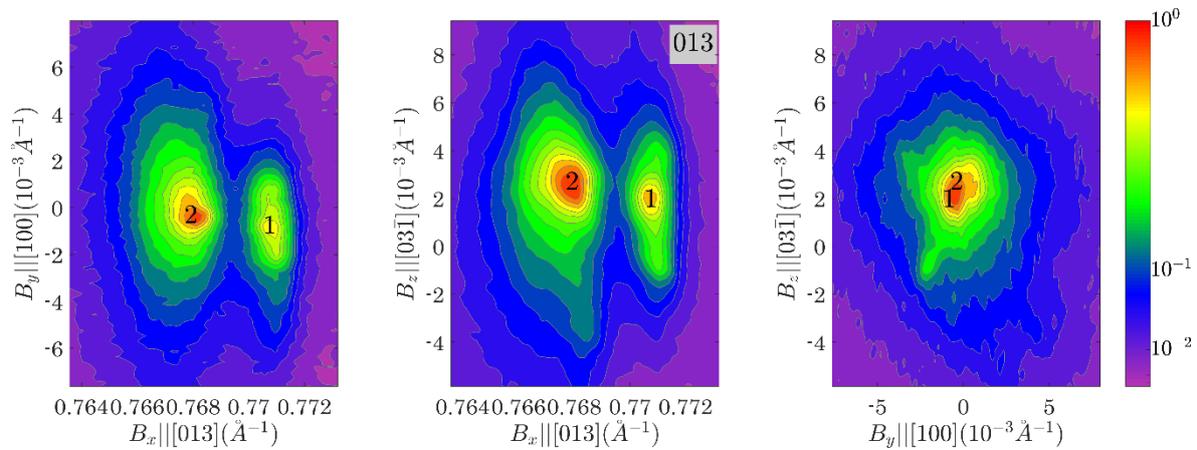

**Figure S7** The same as Figure S1 but for the case of 013 reflections from multi-domain PZT crystal.

**Table S7** The same as Table S1 but for the case of 013 reflection from multi-domain PZT crystal.

| | Sub-peak number | 1 | 2 |
|---|---|---|---|
| | $|B_{obs}|$, Å$^{-1}$ | 0.7708 | 0.7680 |
| | $|B_{calc}|$, Å$^{-1}$ | 0.7708 | 0.7682 |
| | Domain assignments | 3 or 4 | 1 or 2 |
| | $\Delta B_1$ ($10^{-2}$) | -0.2 | -0.1 |
| $\Delta B$ | $\Delta B_2$ ($10^{-2}$) | 0.5 | 0.4 |
| | $\Delta B_3$ ($10^{-2}$) | 1.0 | -0.1 |

**Table S8** The same as Table S2 but for the case of 013 reflections from multi-domain PZT crystal.

| $m$ | $n$ | $\Delta \boldsymbol{B}$ (equation) | $\Delta \boldsymbol{B}$ (calculated), $10^{-2}$ | $\Delta \boldsymbol{B}$ (measured) $10^{-2}$ | Sub peaks pair, $m$ and $n$ (According the numbering in Figure S7) | |
|---|---|---|---|---|---|---|
| $2(A_1)$ | $3(A_2)$ | $\xi(K-H)\begin{pmatrix}0\\0\\1\end{pmatrix}$ | $1.1\begin{pmatrix}0\\0\\1\end{pmatrix}$ | $\begin{pmatrix}-0.1\\+0.1\\+1.1\end{pmatrix}$ | 2 | 1 |
| $1(C)$ | $4(A_3)$ | $\xi(H+K)\begin{pmatrix}0\\0\\1\end{pmatrix}$ | $1.1\begin{pmatrix}0\\0\\1\end{pmatrix}$ | $\begin{pmatrix}-0.1\\+0.1\\+1.1\end{pmatrix}$ | 2 | 1 |

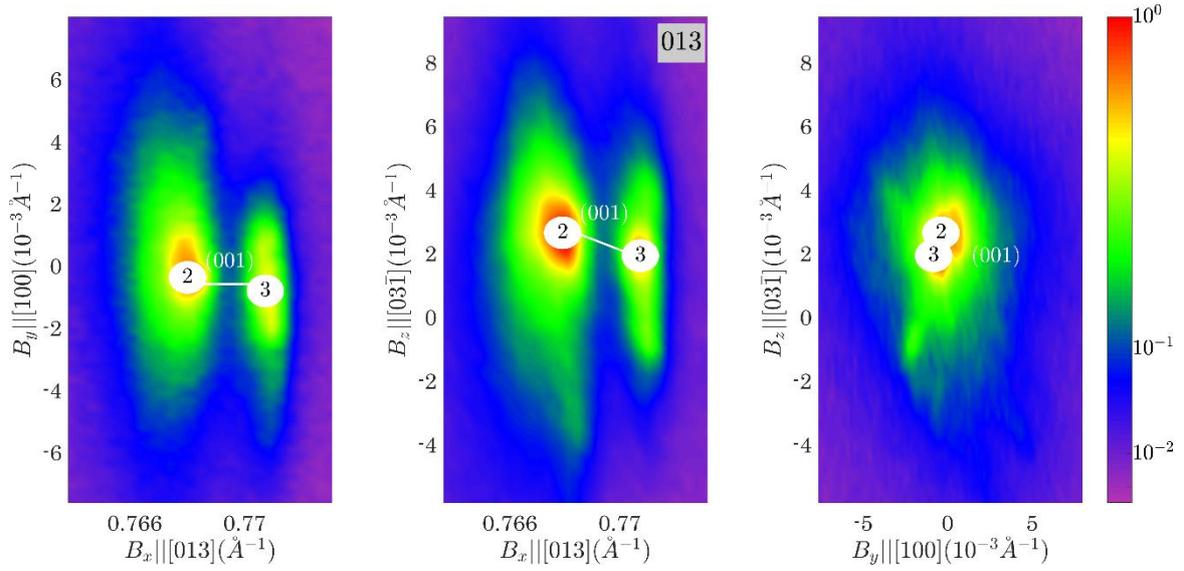

**Figure S8** The same as in the Figure S2 but for the case of 013 family of reflection of rhombohedral PZT.

This example demonstrate the case when ambiguous assignment of sub-peaks to domains is not possible. Any of the two following assignment can be suggested.

- o  The sub-peaks 2 and 1 are assigned to the rhombohedral domains $2(A_1)$ and $3(A_2)$. These domains meet $(001)$ domain wall.
- o  Alternatively, the sub-peaks 2 and 1 are assigned to the rhombohedral domains $1(C)$ and $4(A_3)$, meeting along $(001)$-oriented domain wall too.

**S2.2. Example 2: Splitting of 113 reflection / PZT.**

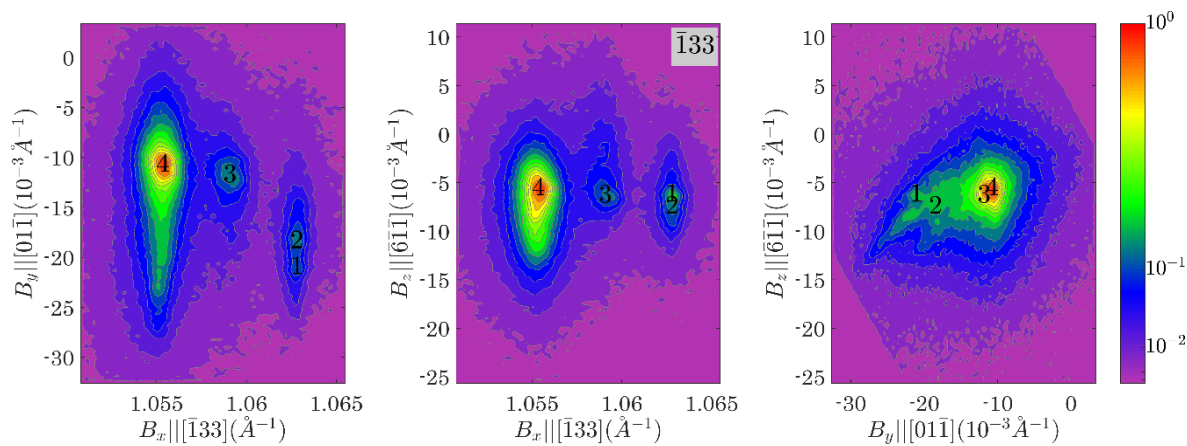

**Figure S9** The same as Figure S1 but for the case of $\bar{1}33$ reflections from multi-domain PZT crystal.

**Table S9** The same as Table S2 but for the case of $\bar{1}33$ family of reflections from multi-domain PZT crystal

| | Sub-peak number | 1 | 2 | 3 | 4 |
|---|---|---|---|---|---|
| | $|B_{obs}|$, Å$^{-1}$ | 1.0630 | 1.0630 | 1.0590 | 1.0553 |
| | $|B_{calc}|$, Å$^{-1}$ | 1.0627 | 1.0627 | 1.0590 | 1.0553 |
| | Domain assignments | 3,4 | 3,4 | 1 | 2 |
| | $\Delta B_1$ (10$^{-2}$) | -0.7 | -0.2 | -0.2 | -0.2 |
| $\Delta B$ | $\Delta B_2$ (10$^{-2}$) | -0.4 | 0.5 | 1.3 | 0.5 |
| | $\Delta B_3$ (10$^{-2}$) | 3.8 | 3.2 | 0.1 | -1.3 |

**Table S10** The same as Table S2 but for the case of $\bar{1}13$ family of reflections of rhombohedral PZT crystal

| m | n | $\Delta B$ (equation) | $\Delta B$ (calculated), 10$^{-2}$ | $\Delta B$ (measured) 10$^{-2}$ | Sub peaks pair (According the numbering in Figure S9) | |
|---|---|---|---|---|---|---|
| 1(C) | 2($A_1$) | $2\eta H \begin{pmatrix} 0 \\ 1 \\ 1 \end{pmatrix}$ | $-1.1 \begin{pmatrix} 0 \\ 1 \\ 1 \end{pmatrix}$ | $\begin{pmatrix} +0.0 \\ -0.8 \\ -1.4 \end{pmatrix}$ | 3 | 4 |

| | | | | | | | |
|---|---|---|---|---|---|---|---|
| 2($A_1$) | 3($A_2$) | $\xi(K-H)\begin{pmatrix}0\\0\\1\end{pmatrix}$ | | $+4.4\begin{pmatrix}0\\0\\1\end{pmatrix}$ | $\begin{pmatrix}+0.0\\+0.0\\+4.5\end{pmatrix}$ | 4 | 2 |

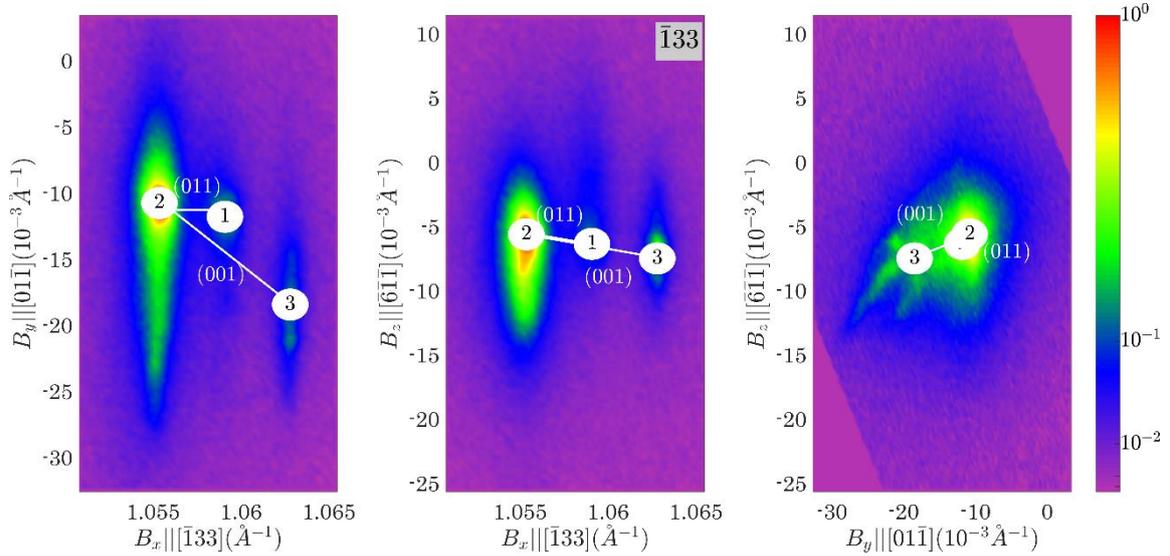

**Figure S10** The same as in the Figure S2 but for the case of $\bar{1}33$ family of reflection of rhombohedral PZT.

According to Table S10 and Figure 10 the following assignment can be made:

- The sub-peaks 3 and 4 are assigned to the rhombohedral domains $1(C)$ and $2(A_1)$. These domains meet along $(011)$ domain wall.
- The sub-peaks 4 and 2 can be assigned to the tetragonal domains $2(A_1)$ and $3(A_2)$. These domains meet along $(001)$ domain wall.

### S2.3. Example 3: Splitting of 233 reflection / PZT

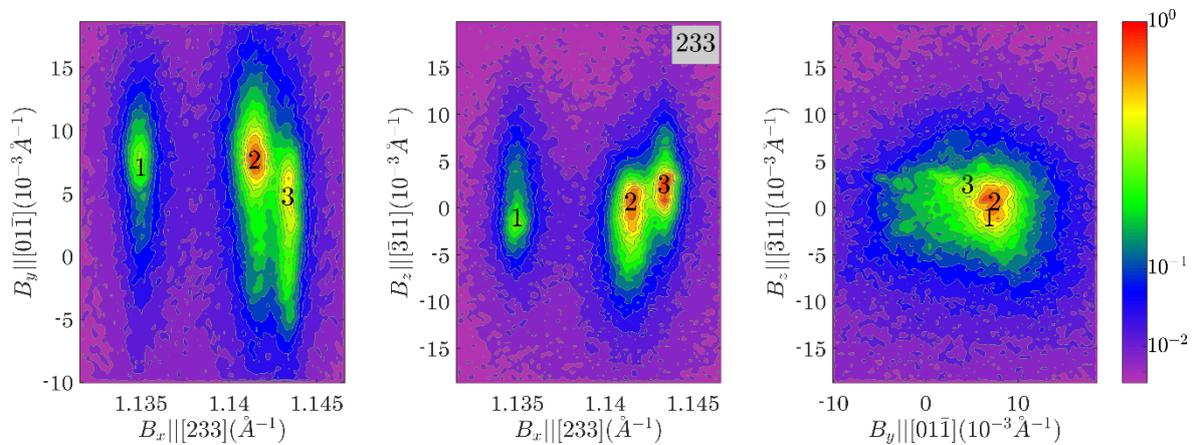

**Figure S11** The same as Figure S1 but for the case of 233 reflections from PZT crystal.

**Table S11** The same as Table S1 but for the case of 233 reflection of multi-domain PZT crystal

| Sub-peak number | | 1 | 2 | 3 |
|---|---|---|---|---|
| $|B_{obs}|$, Å$^{-1}$ | | 1.1350 | 1.1415 | 1.1433 |
| $|B_{calc}|$, Å$^{-1}$ | | 1.1347 | 1.1415 | 1.1433 |
| Domain assignments | | 1 | 2 | 3,4 |
| $\Delta B$ | $\Delta B_1$ ($10^{-2}$) | -1.4 | -0.9 | -1.3 |
| | $\Delta B_2$ ($10^{-2}$) | -1.7 | 0.3 | 0.2 |
| | $\Delta B_3$ ($10^{-2}$) | -2.1 | -0.4 | 1.2 |

**Table S12** The same as Table S2 but for the case of 233 family of reflections of rhombohedral PZT crystal

| $m$ | $n$ | $\Delta B$ (equation) | $\Delta B$ (calculated), $10^{-2}$ | $\Delta B$ (measured) $10^{-2}$ | Sub peaks pair (According the numbering in Figure S11) | |
|---|---|---|---|---|---|---|
| 1(C) | 2($A_1$) | $2\eta H \begin{pmatrix} 0 \\ 1 \\ 1 \end{pmatrix}$ | $2.2 \begin{pmatrix} 0 \\ 1 \\ 1 \end{pmatrix}$ | $\begin{pmatrix} +0.5 \\ +2.0 \\ +1.7 \end{pmatrix}$ | 1 | 2 |
| 2($A_1$) | 3($A_2$) | $\xi(K-H) \begin{pmatrix} 0 \\ 0 \\ 1 \end{pmatrix}$ | $1.1 \begin{pmatrix} 0 \\ 0 \\ 1 \end{pmatrix}$ | $\begin{pmatrix} -0.4 \\ -0.1 \\ +1.6 \end{pmatrix}$ | 2 | 3 |

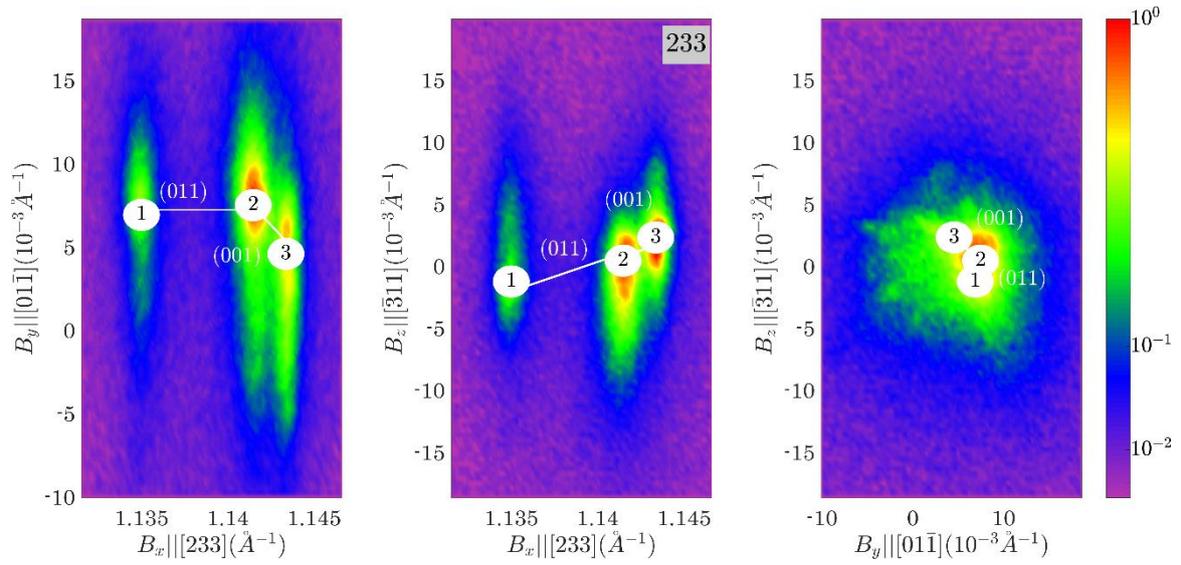

**Figure S12** The same as in the Figure S2 but for the case of 233 family of reflection of rhombohedral PZT.

According to Table S12 and Figure S12 the following assignment can be made:

- The sub-peaks 1 and 2 are assigned to the rhombohedral domains 1(C) and 2($A_1$). These domains meet along (011) domain wall.
- The sub-peaks 2 and 3 can be assigned to the tetragonal domains 2($A_1$) and 3($A_2$). These domains meet along (001) domain wall.